\documentclass[fleqn,usenatbib]{mnras}

\usepackage{subfig}
\usepackage{comment}
\usepackage{longtable}
\usepackage{caption}
\usepackage{threeparttable}
\usepackage{booktabs}
\usepackage[figuresright]{rotating}
\newcommand{\be}{\begin{equation}}
\newcommand{\ee}{\end{equation}}
\newcommand{\bea}{\begin{eqnarray}}
\newcommand{\eea}{\end{eqnarray}}
\newcommand{\hunit}{$\rm{km \ s^{-1} \ Mpc^{-1}}$}
\newcommand{\lcdm}{$\Lambda$CDM}
\newcommand{\pcdm}{$\phi$CDM}
\usepackage{array}
\makeatletter
\newcommand{\thickhline}{%
    \noalign {\ifnum 0=`}\fi \hrule height 1pt
    \futurelet \reserved@a \@xhline
}
\newcolumntype{"}{@{\hskip\tabcolsep\vrule width 1pt\hskip\tabcolsep}}
\makeatother

\newcommand{\hiig}{H\,\textsc{ii}G}
\newcommand{\hii}{H\,\textsc{ii}}
\newcommand{\Om}{\Omega_{\rm m_0}}
\newcommand{\Ok}{\Omega_{\rm k_0}}
\newcommand{\om}{$\Omega_{\rm m_0}$}
\newcommand{\ok}{$\Omega_{\rm k_0}$}

\newcommand{\obh}{\Omega_{\rm b_0}\!h^2}
\newcommand{\och}{\Omega_{\rm c_0}\!h^2}
\newcommand{\obhs}{$\Omega_{\rm b_0}\!h^2$}
\newcommand{\ochs}{$\Omega_{\rm c_0}\!h^2$}

\usepackage[T1]{fontenc}

\DeclareRobustCommand{\VAN}[3]{#2}
\let\VANthebibliography\thebibliography
\def\thebibliography{\DeclareRobustCommand{\VAN}[3]{##3}\VANthebibliography}

\usepackage{graphicx}	
\usepackage{amsmath}	
\usepackage{amssymb}	
\usepackage{fnpct}

\title[SN Ia, BAO, and H(z) constraints]{Using Pantheon and DES supernova, baryon acoustic oscillation, and Hubble parameter data to constrain the Hubble constant, dark energy dynamics, and spatial curvature}

\author[Cao et al.]{
Shulei Cao,$^{1}$\thanks{E-mail: shulei@phys.ksu.edu}
Joseph Ryan,$^{1}$\thanks{E-mail: jwryan@phys.ksu.edu}
Bharat Ratra$^{1}$\thanks{E-mail: ratra@phys.ksu.edu}
\\
$^{1}$Department of Physics, Kansas State University, 116 Cardwell Hall, Manhattan, KS 66502, USA
}

\date{Accepted XXX. Received YYY; in original form ZZZ}

\pubyear{2021}

\begin{document}
\label{firstpage}
\pagerange{\pageref{firstpage}--\pageref{lastpage}}
\maketitle

\begin{abstract}
We use Pantheon Type Ia supernova (SN Ia) apparent magnitude, DES-3yr binned SN Ia apparent magnitude, Hubble parameter, and baryon acoustic oscillation measurements to constrain six spatially flat and non-flat cosmological models. These sets of data provide mutually consistent cosmological constraints in the six cosmological models we study. A joint analysis of these data sets provides model-independent estimates of the Hubble constant, $H_0=68.8\pm1.8\ \rm{km \ s^{-1} \ Mpc^{-1}}$, and the non-relativistic matter density parameter, $\Omega_{\rm m_0}=0.294\pm0.020$. Although the joint constraints prefer mild dark energy dynamics and a little spatial curvature, they do not rule out dark energy being a cosmological constant and flat spatial hypersurfaces. We also add quasar angular size and \hii\ starburst galaxy measurements to the combined data set and find more restrictive constraints.
\end{abstract}




\section{Introduction} \label{sec:intro}

That the Universe is currently in a phase of accelerated expansion is well-supported by observations but not fully explained by fundamental theory (see e.g. \citealp{Ratra_Vogeley,Martin,Coley_Ellis}). The standard spatially flat \lcdm\ model \citep{peeb84} interprets this phenomenon as a consequence of dark energy with negative pressure (a cosmological constant, $\Lambda$) and requires the major part of the energy budget of the Universe to consist of time-independent dark energy and cold dark matter (CDM). Although flat \lcdm\ is consistent with many observations (see e.g. \citealp{Farooq_Ranjeet_Crandall_Ratra_2017,scolnic_et_al_2018,planck2018b,eBOSS_2020}),\footnote{Note that the \textit{Planck} TT,TE,EE+lowE+lensing data favor positive spatial curvature \citep{planck2018b} but are consistent with a spatially flat model at $1.63\sigma$.} there exist some potential observational discrepancies \citep{riess_2019, martinelli_tutusaus_2019} and theoretical puzzles (e.g., \citealp{Martin}), which leaves room for other cosmological models, including non-flat \lcdm. As the quality and quantity of observational data grow in time, constraining these models is within reach. Many workers have investigated the merits of the flat and non-flat XCDM parametrizations and \pcdm\ models, where dark energy dynamics and spatial curvature come into play.\footnote{For observational constraints on spatial curvature see \cite{Farooq_Mania_Ratra_2015}, \cite{Chen_et_al_2016}, \cite{rana_jain_mahajan_mukherjee_2017}, \cite{ooba_etal_2018a, ooba_etal_2018b, ooba_etal_2018c}, \cite{yu_etal_2018}, \cite{park_ratra_2018, park_ratra_2019a, park_ratra_2019c, park_ratra_2020}, \cite{wei_2018}, \cite{DES_2019}, \cite{handley_2019a}, \cite{jesus_etal_2019}, \cite{li_etal_2020}, \cite{geng_etal_2020}, \cite{kumar_etal_2020}, \cite{efstathiou_gratton_2020}, \cite{divalentino_etal_2020}, \cite{divalentino_etal_2020b}, \cite{gao_etal_2020}, \cite{Abbassi_2020}, \cite{Yang_2020}, \cite{Agudelo_Ruiz_2020}, \cite{Velasquez-Toribio_2020}, \cite{Vagnozzi_2020a,Vagnozzi_2020b}, and references therein.}\footnote{For observational constraints on the \pcdm\ model see \cite{yashar_et_al_2009}, \cite{Samushia_2010}, \cite{Campanelli_etal_2012}, \cite{Avsajanishvili_2015}, \cite{Sola_etal_2017}, \cite{Sola_perez_gomez_2018,sola_gomez_perez_2019}, \cite{zhai_blanton_slosar_tinker_2017}, \cite{ooba_etal_2018b,ooba_etal_2019}, \cite{sangwan_tripathi_jassal_2018}, \cite{singh_etal_2019}, \cite{Khadka_2020a,Khadka_2020b,Khadka_2020c,Khadka_2020d}, \cite{Urena-Lopez_2020}, and references therein.}

Many observational data sets have been used to place constraints on the parameters of cosmological models, such as the equation of state parameter ($w$) of dark energy. Most recently, in \cite{Caoetal_2020b}, we used Hubble parameter ($H(z)$), baryon acoustic oscillation (BAO), quasar angular size (QSO), quasar X-ray and UV flux, \hii\ starburst galaxy (\hiig), and gamma-ray burst (GRB) data to constrain this parameter (among others). The tightest constraints on $w$, we found, come from low-redshift $H(z)$ (cosmic chronometer) and BAO (standard ruler) data, with the standard candle data (\hiig\ and GRB) giving very broad constraints. In this paper we combine measurements of the distances to 1255 Type Ia supernovae (SNe Ia) with our set of $H(z)$ and BAO data (along with QSO and \hiig\ observations) to obtain tight cosmological parameter constraints.

The usefulness of SN Ia data to cosmology is well-known. SN Ia measurements revealed the accelerated expansion of the Universe over twenty years ago, and they are employed today to place constraints on cosmological parameters and to break parameter degeneracies. Over this time period, the sample size of SN Ia distance measurements has grown considerably, and the analysis and mitigation of systematic uncertainties has improved \citep{DES_2019c, DES_2019d}. Supernovae are therefore a reasonably empirically well-understood cosmological probe\footnote{Though the relatively simpler physics underlying cosmic microwave background (CMB) anisotropies and BAO makes those probes better understood than SNe Ia.}, and so can be used to obtain reliable constraints on cosmological model parameters.

In our earlier studies that made use of BAO data (e.g., \citealp{Ryan_2, Caoetal_2020b}), we relied on CMB-derived values of the baryon density\footnote{Here $\Omega_{\rm b_0}$ is the baryon density parameter and $h=H_0/(100\ \rm{km \ s^{-1} \ Mpc^{-1}})$.} $\obh$ in order to compute the size of the sound horizon $r_{s}$. The size of the sound horizon is needed to calibrate the BAO scale (see Table \ref{tab:BAO}), so the constraints we derived from our BAO measurements were indirectly dependent on CMB physics. \cite{park_ratra_2018, park_ratra_2019a, park_ratra_2019c} computed $\obh$ within each of the six models we study (namely flat/non-flat \lcdm, flat/non-flat XCDM, and flat/non-flat \pcdm) from CMB data using primordial energy density fluctuation power spectra $P(k)$ appropriate for flat and curved geometries \citep{Lucchin_1985, ratra_1989,ratra_2017,ratra_peebles_1995}. Other power spectra have been considered in the non-flat case \citep{Lesgourgues_2014,Bonga_2016,handley_2019b,Thavanesan_2021}. Since we do not make use of $P(k)$, the controversy associated with $P(k)$ in non-flat models is avoided in our analyses here.

The constraints from $H(z)$ + BAO data and from SN Ia data are not inconsistent, and so these data can be jointly used to constrain cosmological parameters. \cite{park_ratra_2019b} used $H(z)$, BAO, and Pantheon SN Ia apparent magnitude (SN-Pantheon) measurements in such a joint analysis. Here we use a more recent BAO data compilation and new DES-3yr binned SN Ia apparent magnitude (SN-DES) data. We find for all combinations of data we study here that all or almost all of the favored parameter space corresponds to currently accelerating cosmological expansion. The most reliable constraints come from the $H(z)$ + BAO + SN-Pantheon + SN-DES (HzBSNPD) data combination, with fairly model-independent determinations of the Hubble constant, $H_0=68.8\pm1.8\ \rm{km \ s^{-1} \ Mpc^{-1}}$, and the non-relativistic matter density parameter, $\Omega_{\rm m_0}=0.294\pm0.020$. The estimate of $H_0$ is in better agreement with the median statistics $H_0 = 68 \pm 2.8$ \hunit\ estimate of \cite{chenratmed} and the \cite{planck2018b} estimate of $H_0 = 67.4 \pm 0.5$ \hunit\ than with the local $H_0 = 74.03 \pm 1.42$ \hunit\ measurement of \cite{riess_etal_2019}. The combined measurements are consistent with the spatially flat \lcdm\ model, but also favor some dark energy dynamics, as well as a little non-zero spatial curvature energy density. More restrictive constraints are derived when these data are combined with QSO and \hiig\ data.

This paper is organized as follows. Section \ref{sec:model} summarizes the models we analyze. In Section \ref{sec:data} the data used are introduced and our method of analyzing these data is described in Section \ref{sec:analysis}. We present our results in Section \ref{sec:results}, and our conclusions in Section \ref{sec:conclusion}.

\section{Cosmological models}
\label{sec:model}

We seek to obtain constraints on the parameters of the flat and non-flat \lcdm, XCDM, and \pcdm\ models and to compare how well these models fit the observations we study. These models have been described in \cite{Ryan_2} and \cite{Caoetal_2020b}; see those papers for more details. Our approach here differs from that of those earlier papers in that, instead of varying the non-relativistic matter density parameter \om\ as a free parameter, we vary the baryonic (\obhs) and cold dark matter (\ochs) densities as free parameters, treating \om\ as a derived parameter.\footnote{We do this to eliminate the dependence of the BAO data on CMB physics; see Section \ref{sec:data} for details.} 

The expansion rate function $E(z) \equiv H(z)/H_0$ takes the following form in the non-flat \lcdm\ model:
\begin{equation}
\label{eq:E(z)_NFLCDM}
    E(z) = \sqrt{\Om\left(1 + z\right)^3 + \Ok\left(1 + z\right)^2 + \Omega_{\Lambda}},
\end{equation}
where $z$ is the redshift,
\begin{equation}
\label{eq:Om}
    \Om = \frac{\obh\ + \och}{h^2} + \Omega_{\nu_0},
\end{equation}
and
\begin{equation}
    \Omega_{\Lambda} = 1 - \Om - \Ok.
\end{equation}
$\Ok$ is the curvature energy density parameter, and $\Omega_{\nu_0}$ is the neutrino energy density parameter, which we, following \cite{Carter_2018}, set to $\Omega_{\nu_0} = 0.0014$ for all models. The non-flat \lcdm\ model therefore has four free parameters: $h$, $\obh$\!, $\och$\!, and \ok. The flat \lcdm\ model is a special case with $\Ok = 0$.

In the non-flat XCDM parametrization, the expansion rate function takes the form
\begin{equation}
    \label{eq:E(z)_NFXCDM}
    E(z) = \sqrt{\Om\left(1 + z\right)^3 + \Ok\left(1 + z\right)^2 + \Omega_{\rm X_0}\left(1 + z\right)^{3\left(1 + w_{\rm X}\right)}},
\end{equation}
where
\begin{equation}
    \Omega_{\rm X_0} = 1 - \Om - \Ok.
\end{equation}
Here $w_{\rm X}$ is the equation of state parameter of the X-fluid. The non-flat XCDM parametrization therefore has five free parameters: $h$, $\obh$\!, $\och$\!, $\Ok$, and $w_{\rm X}$. The flat XCDM parametrization is a special case in which $\Ok = 0$.

In the non-flat \pcdm\ model \citep{peebrat88,ratpeeb88,pavlov13}, a scalar field $\phi$ plays the role of a time-varying cosmological ``constant''. The expansion rate function in this model takes the form
\begin{equation}
    E(z) = \sqrt{\Om\left(1 + z\right)^3 + \Ok\left(1 + z\right)^2 + \Omega_{\phi}(z,\alpha)},
\end{equation}
where the energy density parameter of the scalar field $\phi$, $\Omega_{\phi}(z,\alpha)$, is determined by numerically integrating the scalar field's equations of motion. In this quantity $\alpha$ is the parameter that controls the shape of the inverse power law potential energy density $V(\phi)$ of $\phi$.\footnote{The details of this model are described in \cite{Caoetal_2020a,Caoetal_2020b}.} The non-flat \pcdm\ model therefore has five free parameters: $h$, $\obh$\!, $\och$\!, $\Ok$, and $\alpha$. The flat \pcdm\ model is a special case in which $\Ok = 0$.

\section{Data}
\label{sec:data}

In this paper, we use a combination of $H(z)$, BAO, SN-Pantheon, SN-DES, QSO, and \hiig\ data to constrain the cosmological models we study. 

The $H(z)$ data, compiled in Table 2 of \cite{Ryan_1}, consist of 31 measurements spanning the redshift range $0.070 \leq z \leq 1.965$. The BAO data, which have been updated relative to \cite{Caoetal_2020a}, consist of 11 measurements spanning the redshift range $0.38 \leq z \leq 2.334$, listed in Table \ref{tab:BAO}. 

The SN-Pantheon data, compiled by \cite{scolnic_et_al_2018}, consist of 1048 SN Ia measurements spanning the redshift range $0.01<z<2.3$. The SN-DES data, compiled by \cite{DES_2019d}, consist of 20 binned measurements of 207 SN Ia measurements spanning the redshift range $0.015 \leq z \leq 0.7026$.

The QSO data, listed in Table 1 of \cite{Cao_et_al2017b}, consist of 120 measurements of the angular size
\begin{equation}
    \theta(z) = \frac{l_{\rm m}}{D_{A}(z)},
\end{equation}
spanning the redshift range $0.462 \leq z \leq 2.73$. $l_{\rm m}$ is the characteristic linear size of the quasars in the sample. This quantity is determined by using the Gaussian Process method to reconstruct the expansion history of the Universe from 24 cosmic chronometer measurements over $z < 1.2$. This $H(z)$ function is used to reconstruct the angular size distance $D_{A}(z)$, which can then be used to compute $l_{\rm m}$ given measurements $(\theta_{\rm obs}(z)$) of quasar angular sizes. QSO and $H(z)$ data are therefore somewhat correlated, but the error bars on the constraints derived from QSO data are so large that we do not believe this correlation to be an issue. 

The \hiig\ data consist of 107 low redshift ($0.0088 \leq z \leq 0.16417$) measurements, used in \cite{Chavez_2014} (recalibrated by \citealp{G-M_2019}), and 46 high redshift ($0.636427 \leq z \leq 2.42935$) measurements.

\begin{table}
\centering
\scriptsize
\begin{threeparttable}
\caption{BAO data.}\label{tab:BAO}
\setlength{\tabcolsep}{0.8mm}{
\begin{tabular}{lccc}
\toprule
$z$ & Measurement\tnote{a} & Value & Ref.\\
\midrule
$0.38$ & $D_M\left(r_{s,{\rm fid}}/r_s\right)$ & 1512.39 & \cite{Alam_et_al_2017}\tnote{b}\\
$0.38$ & $H(z)\left(r_s/r_{s,{\rm fid}}\right)$ & 81.2087 & \cite{Alam_et_al_2017}\tnote{b}\\
$0.51$ & $D_M\left(r_{s,{\rm fid}}/r_s\right)$ & 1975.22 & \cite{Alam_et_al_2017}\tnote{b}\\
$0.51$ & $H(z)\left(r_s/r_{s,{\rm fid}}\right)$ & 90.9029 & \cite{Alam_et_al_2017}\tnote{b}\\
$0.61$ & $D_M\left(r_{s,{\rm fid}}/r_s\right)$ & 2306.68 & \cite{Alam_et_al_2017}\tnote{b}\\
$0.61$ & $H(z)\left(r_s/r_{s,{\rm fid}}\right)$ & 98.9647 & \cite{Alam_et_al_2017}\tnote{b}\\
$0.122$ & $D_V\left(r_{s,{\rm fid}}/r_s\right)$ & $539\pm17$ & \cite{Carter_2018}\\
$0.81$ & $D_A/r_s$ & $10.75\pm0.43$ & \cite{DES_2019b}\\
$1.52$ & $D_V\left(r_{s,{\rm fid}}/r_s\right)$ & $3843\pm147$ & \cite{3}\\
$2.334$ & $D_M/r_s$ & 37.5 & \cite{duMas2020}\tnote{c}\\
$2.334$ & $D_H/r_s$ & 8.99 & \cite{duMas2020}\tnote{c}\\
\bottomrule
\end{tabular}}
\begin{tablenotes}[flushleft]
\item[a] $D_M$, $D_V$, $r_s$, $r_{s, {\rm fid}}$, $D_A$, and $D_M$ have units of Mpc, while $H(z)$ has units of \hunit.
\item[b] The six measurements from \cite{Alam_et_al_2017} are correlated; see equation (20) of \cite{Ryan_2} for their correlation matrix.
\item[c] The two measurements from \cite{duMas2020} are correlated; see equation \eqref{CovM1} below for their correlation matrix.
\end{tablenotes}
\end{threeparttable}
\end{table}

The covariance matrix $\textbf{C}$ for the BAO data, taken from \cite{Alam_et_al_2017}, is given in equation (20) of \cite{Ryan_2}. For the BAO data from \cite{duMas2020}, the covariance matrix is
\be
\label{CovM1}
    \textbf{C}=
    \begin{bmatrix}
    1.3225 & -0.1009 \\
    -0.1009 & 0.0380
    \end{bmatrix}.
\ee
The scale of BAO measurements is set by the sound horizon ($r_{s}$) during the epoch of radiation drag. To compute this quantity, we use the approximate formula \citep{PhysRevD.92.123516}
\be
\label{eq:sh}
    r_s=\frac{55.154\exp{[-72.3(\Omega_{\rm \nu_0}h^2+0.0006)^2]}}{(\Omega_{\rm b_0}\!h^2)^{0.12807}(\och+\obh)^{0.25351}} \mathrm{Mpc}.
\ee
In our previous studies we did not vary \obhs\ as a free parameter. Instead we used CMB-derived, model-dependent values of \obhs\ to compute $r_{s}$. Because we vary \obhs\ as a free parameter in this paper, our computations of the sound horizon (and therefore our calibration of the scale of our BAO measurements) are fully independent of CMB physics (at the cost of enlarging the parameter space and so somewhat weakening the constraints).

Following \cite{Conley_et_al_2011} and \cite{Deng_Wei_2018}, we define the theoretical magnitude of a supernova to be
\begin{equation}
\label{eq:m_th}
    m_{\rm th} = 5\log\mathcal{D}_{L}(z) + \mathcal{M},
\end{equation}
where $\mathcal{M}$ is a nuisance parameter to be marginalized over, and $\mathcal{D}_{L}(z)$ is
\begin{equation}
\label{eq:D_L}
    \mathcal{D}_{L}(z) \equiv \left(1 + z_{\rm hel}\right) \int_0^{z_{\rm cmb}} \frac{d\tilde{z}}{E\left(\tilde{z}\right)}.
\end{equation}
In this equation, $z_{\rm hel}$ is the heliocentric redshift, and $z_{\rm cmb}$ is the CMB-frame redshift. In \cite{Conley_et_al_2011}, equation \eqref{eq:D_L} is called the ``Hubble-constant free luminosity distance'', because $E(z)$ does not contain $H_0$. In our case, because we use $h$, \obhs\!, and \ochs\ as free parameters, our expansion rate function (and thus our luminosity distance) depends on the Hubble constant. We therefore obtain weak constraints on $H_0$ from the supernova data, unlike \cite{Conley_et_al_2011} and \cite{Deng_Wei_2018} (see Section \ref{sec:results}, below).

\section{Data Analysis Methodology}
\label{sec:analysis}

We use the \textsc{python} module \textsc{emcee} \citep{2013PASP..125..306F} to maximize the likelihood functions, thereby determining the constraints on the free parameters. In our analyses here the priors on the cosmological parameters are different from zero (and flat) over the ranges $0.005 \leq \Omega_{\rm b_0}\!h^2 \leq 0.1$, $0.001 \leq \Omega_{\rm c_0}\!h^2 \leq 0.99$, $0.2 \leq h \leq 1.0$, $-3 \leq w_{\rm X} \leq 0.2$, $-0.7 \leq \Omega_{\rm k_0} \leq 0.7$, and $0 < \alpha \leq 10$. \om\ is a derived parameter and depends on $h$.

The likelihood functions of $H(z)$, BAO, \hiig, and QSO data are described in \cite{Caoetal_2020a} and \cite{Caoetal_2020b}. For the SN Ia (SN-Pantheon and SN-DES) data, the likelihood function is
\be
\label{eq:LH_SN}
    \mathcal{L}_{\rm SN}= e^{-\chi^2_{\rm SN}/2},
\ee
where, as in \cite{park_ratra_2019b}, $\chi^2_{\rm SN}$ takes the form of equation (C1) in Appendix C of \cite{Conley_et_al_2011} with $\mathcal{M}$ being marginalized. The covariance matrices of the SN Ia data, $\textbf{C}_{\rm SN}$ are the sum of the diagonal statistical uncertainty covariance matrices, $\textbf{C}_{\rm stat} = \rm diag(\sigma^2_{\rm SN})$, and the systematic uncertainty covariance matrices, $\textbf{C}_{\rm sys}$: $\textbf{C}_{\rm SN} = \textbf{C}_{\rm stat} + \textbf{C}_{\rm sys}$.\footnote{Note that the covariance matrices for the SN-DES data are the ones described in equation (18) of \cite{DES_2019d}.} $\sigma_{\rm SN}$ are the SN Ia statistical uncertainties.

As in \cite{Caoetal_2020b}, we use the Akaike Information Criterion ($AIC$) and the Bayesian Information Criterion ($BIC$) to compare the quality of models with different numbers of parameters, where
\be
\label{AIC}
    AIC=-2\ln \mathcal{L}_{\rm max} + 2n,
\ee
and
\be
\label{BIC}
    BIC=-2\ln \mathcal{L}_{\rm max} + n\ln N.
\ee
In the preceding equations, $\mathcal{L}_{\rm max}$, $n$, and $N$ are the maximum value of the considered likelihood function, the number of free parameters in the given model, and the number of used data points (e.g., for SN-Pantheon $N=1048$), respectively.

\section{Results}
\label{sec:results}

\begin{figure*}
\centering
 \subfloat[]{%
    \includegraphics[width=3.5in,height=3.5in]{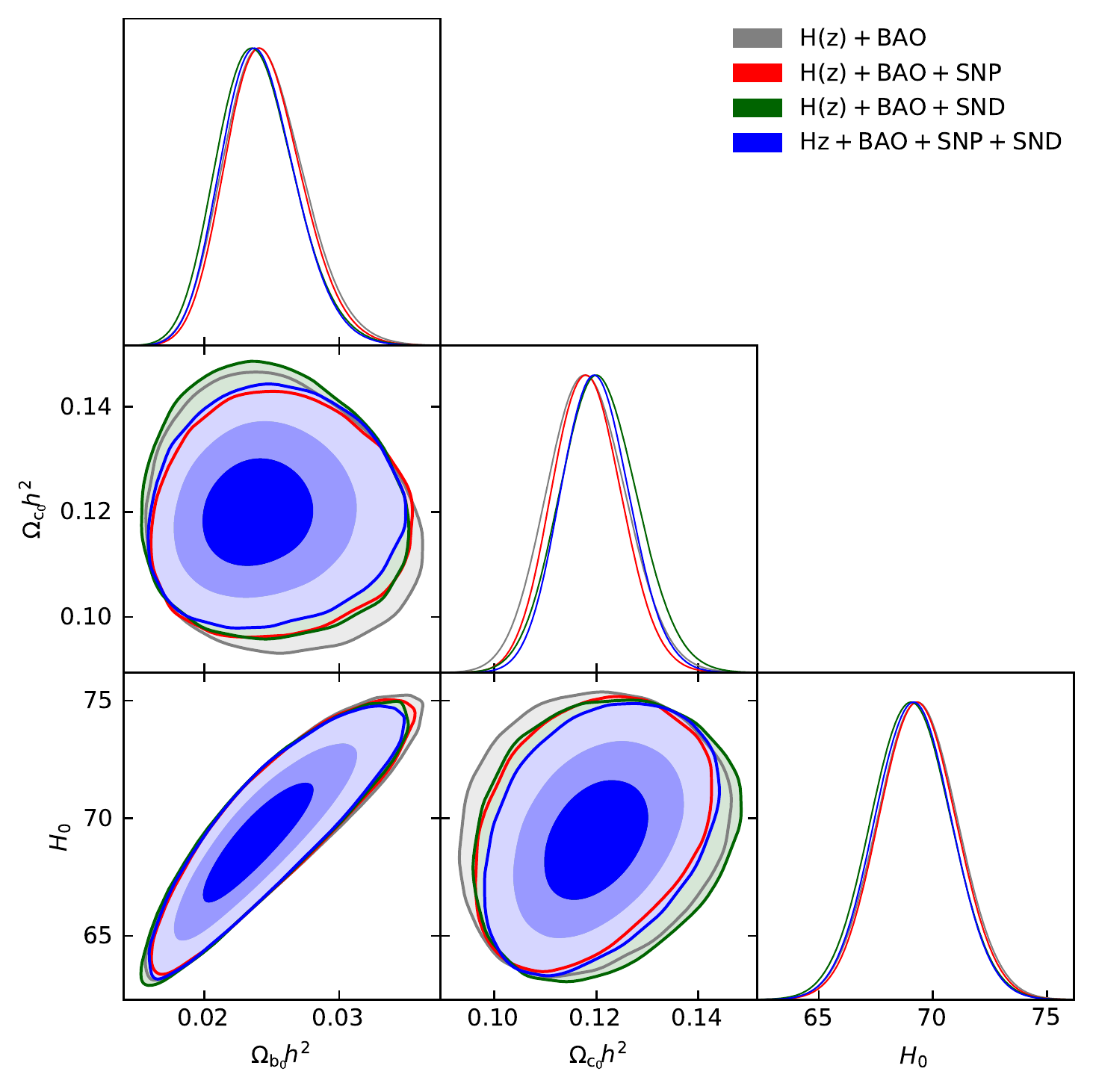}}
 \subfloat[]{%
    \includegraphics[width=3.5in,height=3.5in]{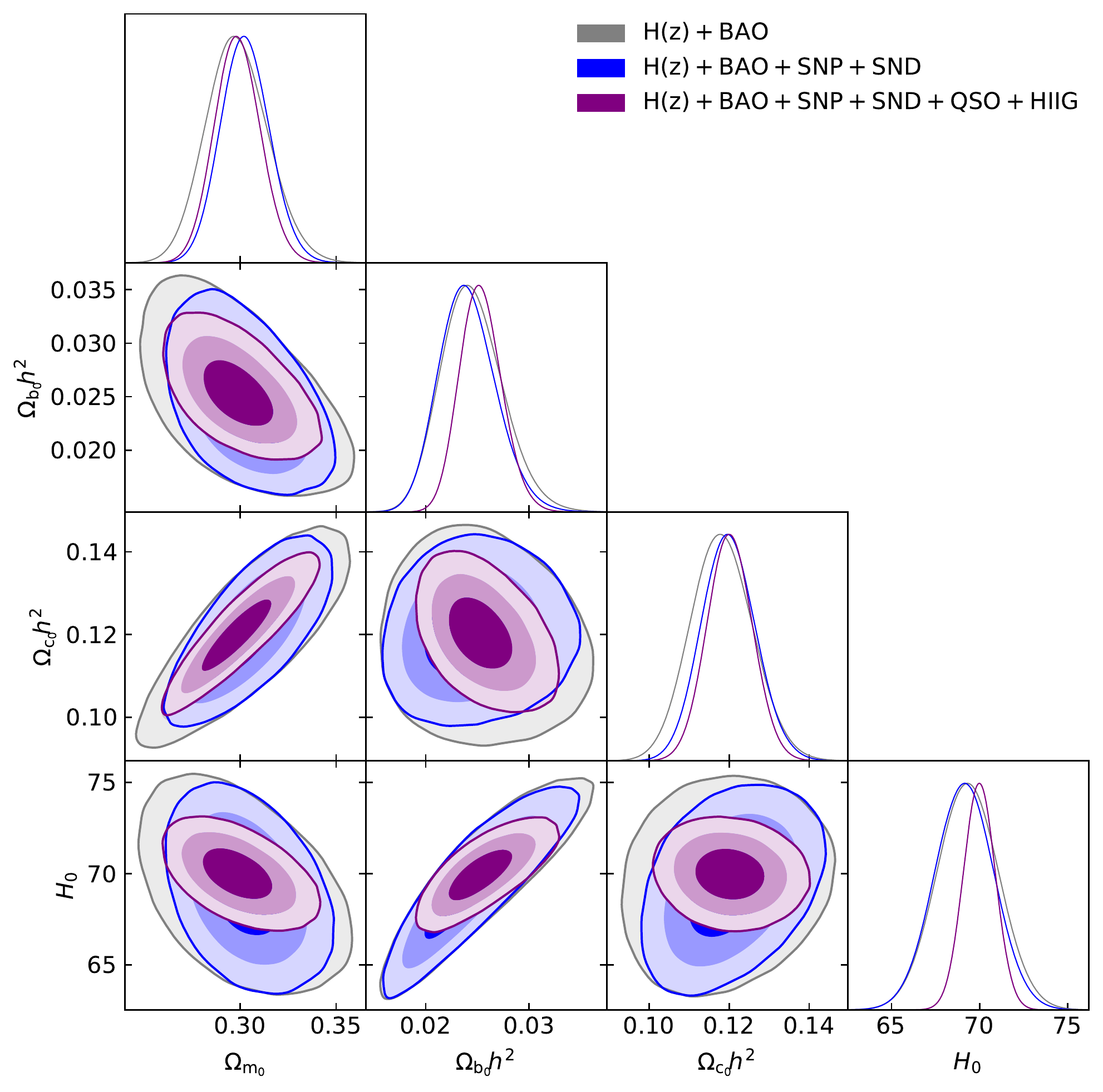}}\\
\caption{1$\sigma$, 2$\sigma$, and 3$\sigma$ confidence contours for flat \lcdm, where the right panel is the comparison including derived cosmological matter density parameter \om. In all cases, the favored parameter space is associated with currently-accelerating cosmological expansion.}
\label{fig1}
\end{figure*}

\begin{figure*}
\centering
 \subfloat[]{%
    \includegraphics[width=3.5in,height=3.5in]{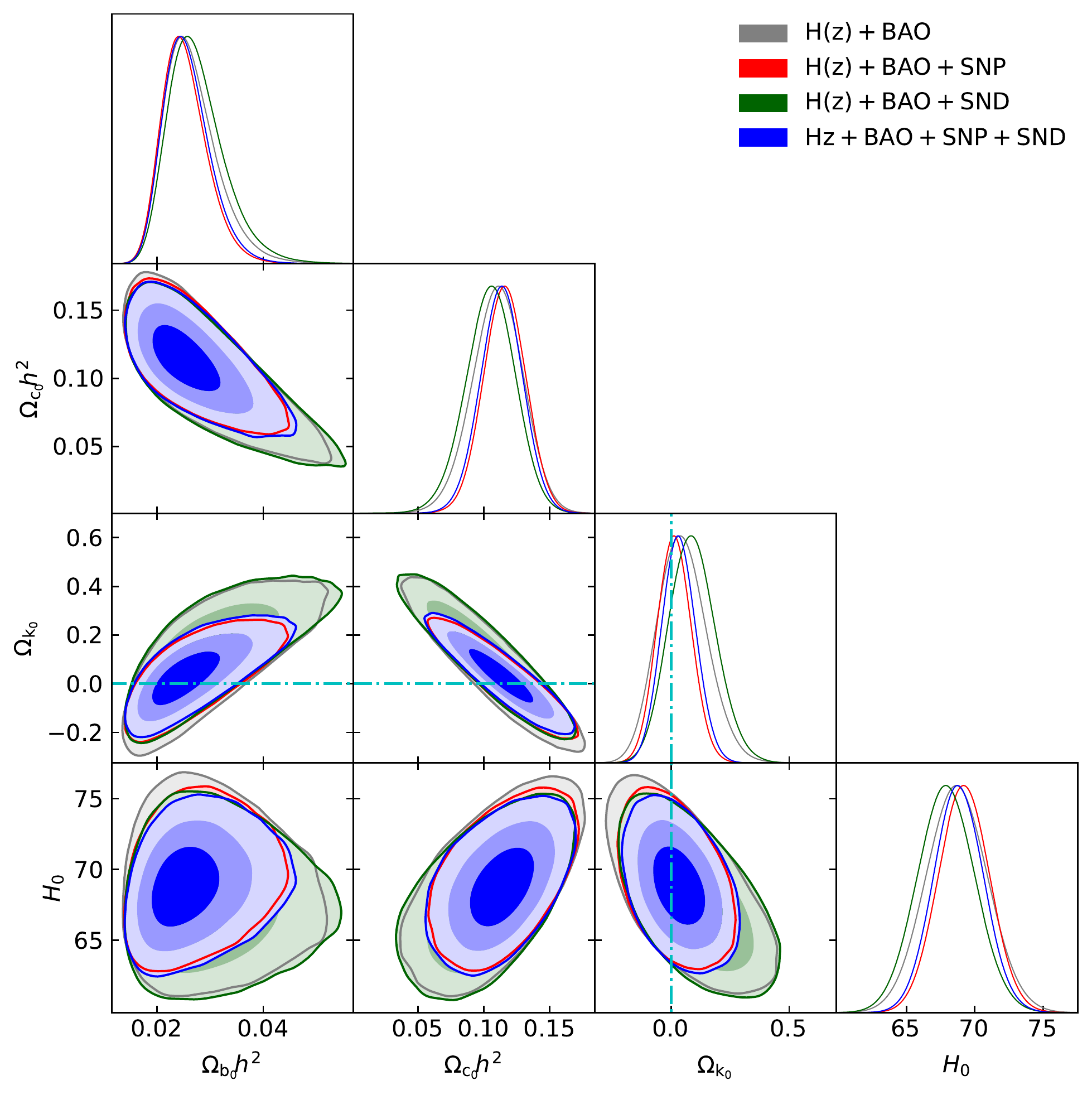}}
 \subfloat[]{%
    \includegraphics[width=3.5in,height=3.5in]{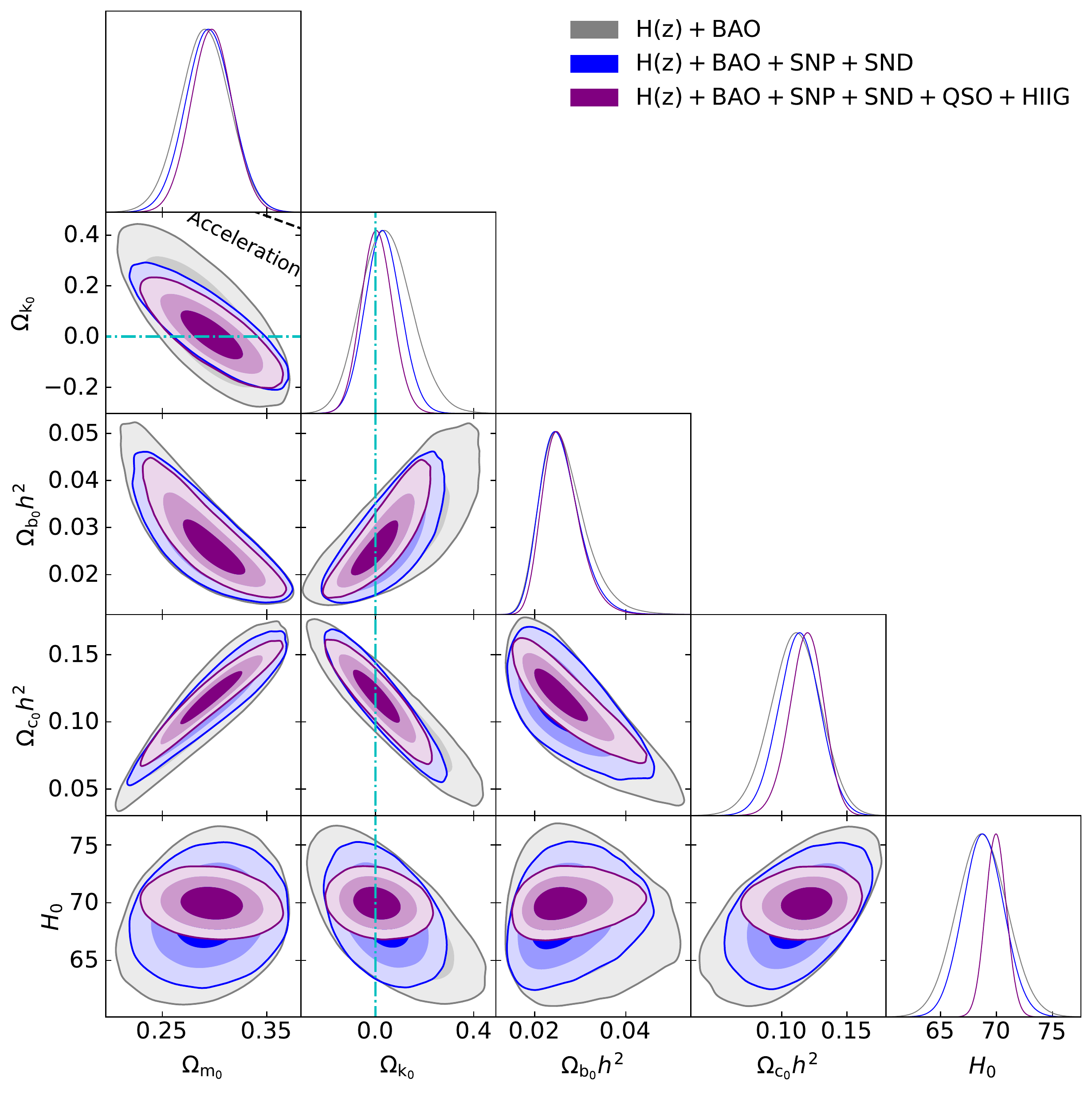}}\\
\caption{Same as Fig. \ref{fig1} but for non-flat \lcdm, where the cyan dash-dot lines represent the flat \lcdm\ case, with closed spatial hypersurfaces either below or to the left. The black dotted line in the right subpanel is the zero-acceleration line, which divides the parameter space into regions associated with currently-accelerating (below left) and currently-decelerating (above right) cosmological expansion. In all cases, the favored parameter space is associated with currently-accelerating cosmological expansion.}
\label{fig2}
\end{figure*}

\begin{figure*}
\centering
 \subfloat[]{%
    \includegraphics[width=3.5in,height=3.5in]{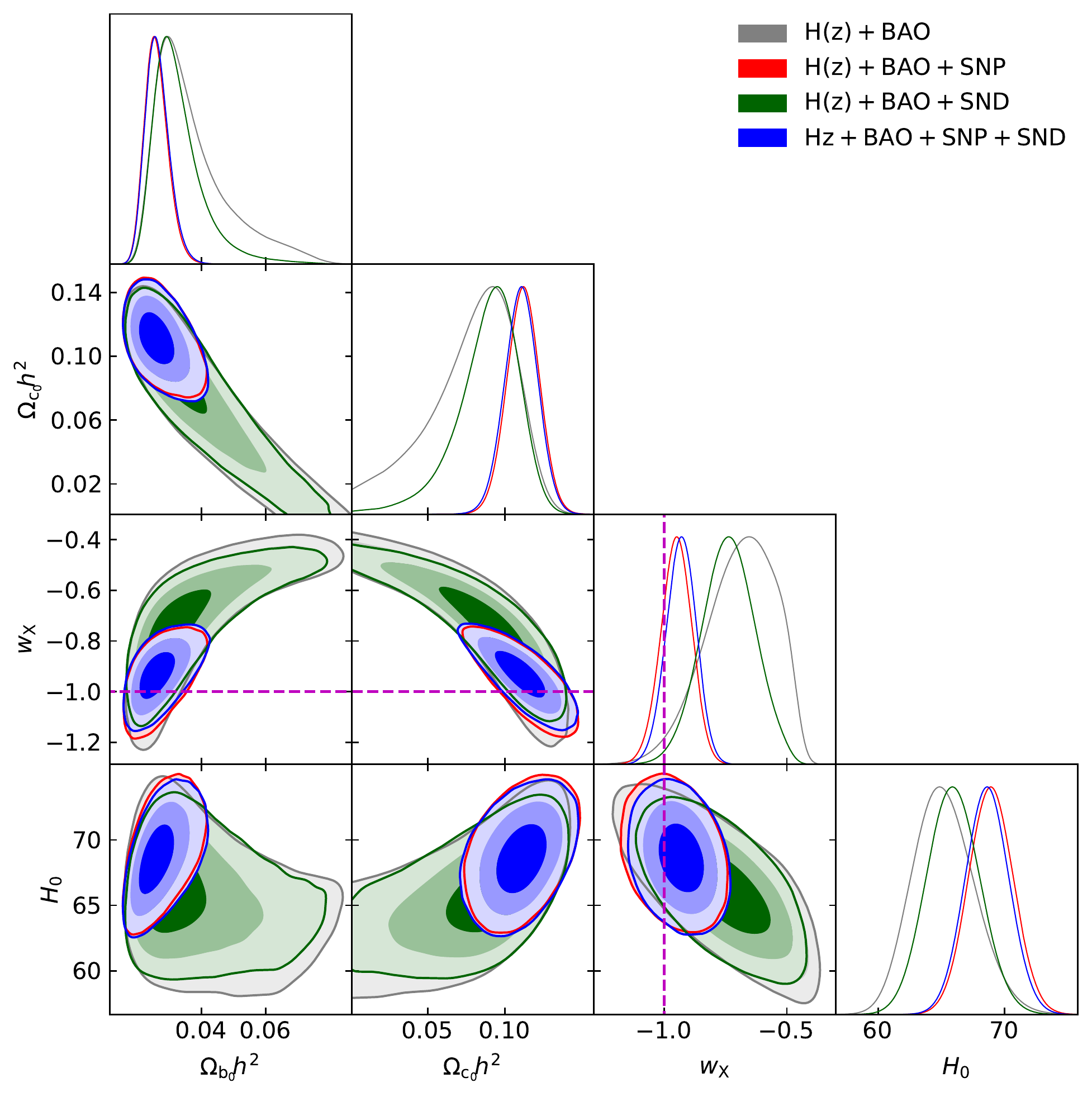}}
 \subfloat[]{%
    \includegraphics[width=3.5in,height=3.5in]{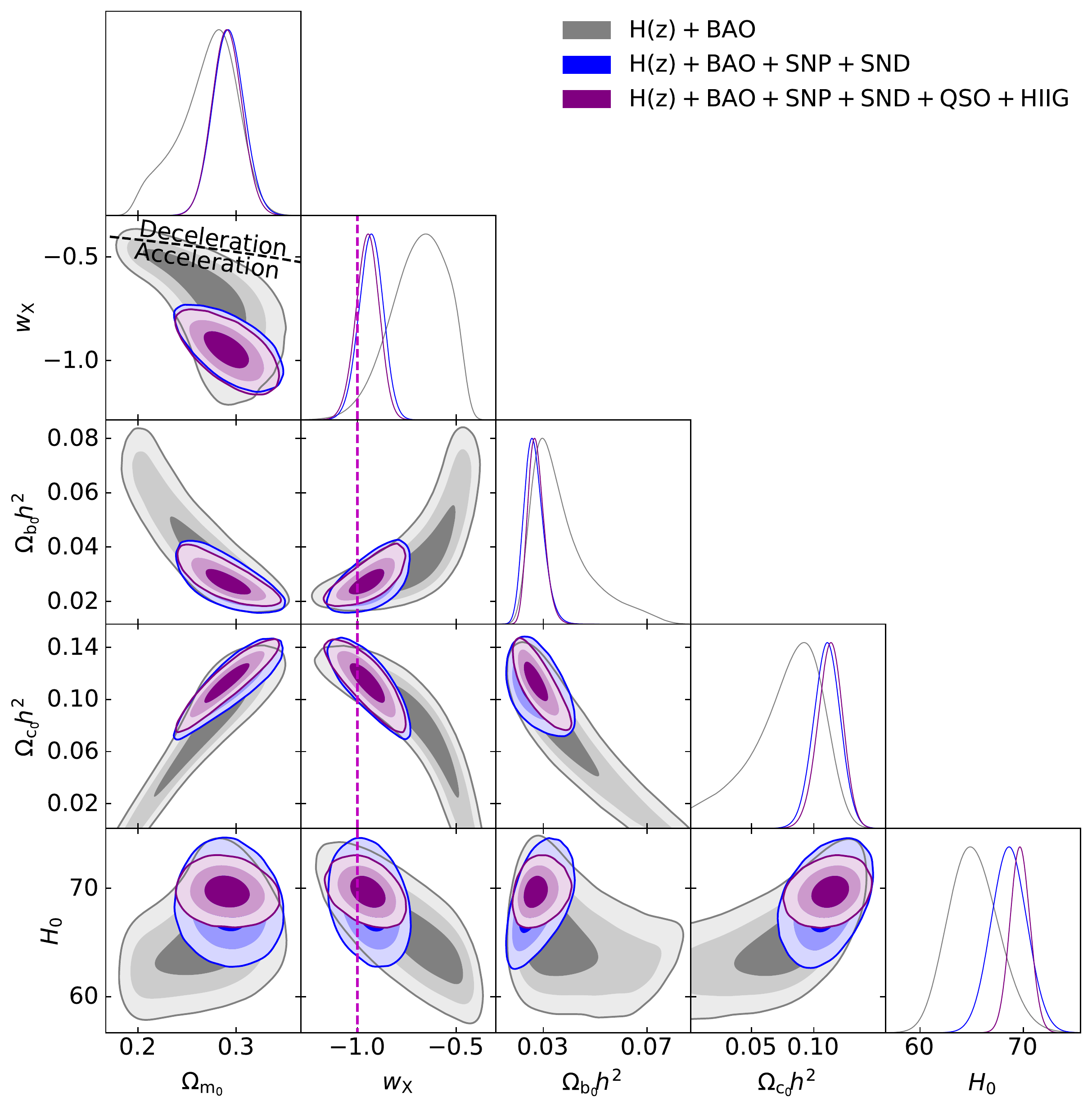}}\\
\caption{1$\sigma$, 2$\sigma$, and 3$\sigma$ confidence contours for flat XCDM. The black dotted line in the right panel is the zero-acceleration line, which divides the parameter space into regions associated with currently-accelerating (below) and currently-decelerating (above) cosmological expansion. In all cases, almost all of the favored parameter space is associated with currently-accelerating cosmological expansion. The magenta lines denote $w_{\rm X}=-1$, i.e. the flat \lcdm\ model.}
\label{fig3}
\end{figure*}

\begin{figure*}
\centering
 \subfloat[]{%
    \includegraphics[width=3.5in,height=3.5in]{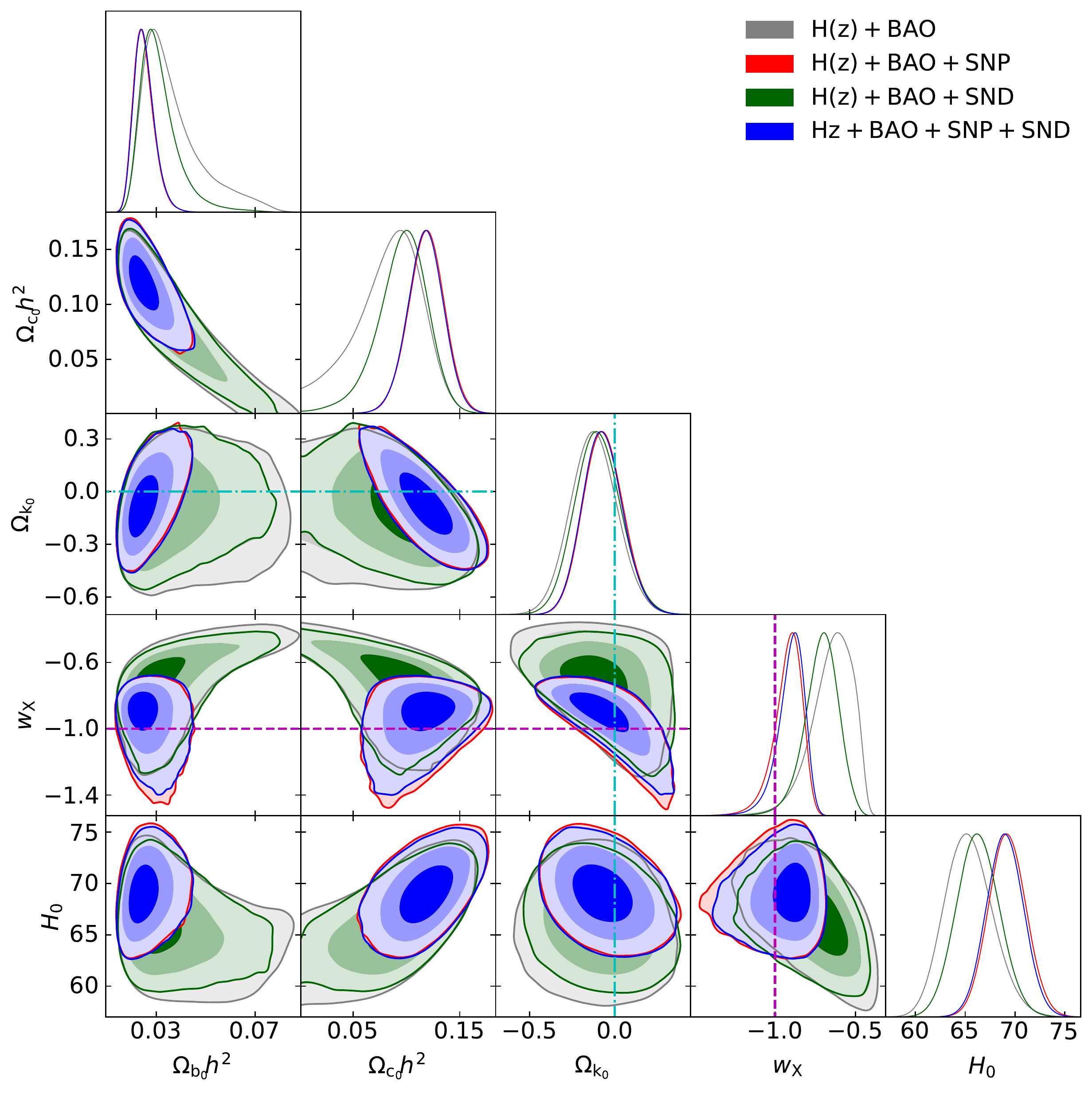}}
 \subfloat[]{%
    \includegraphics[width=3.5in,height=3.5in]{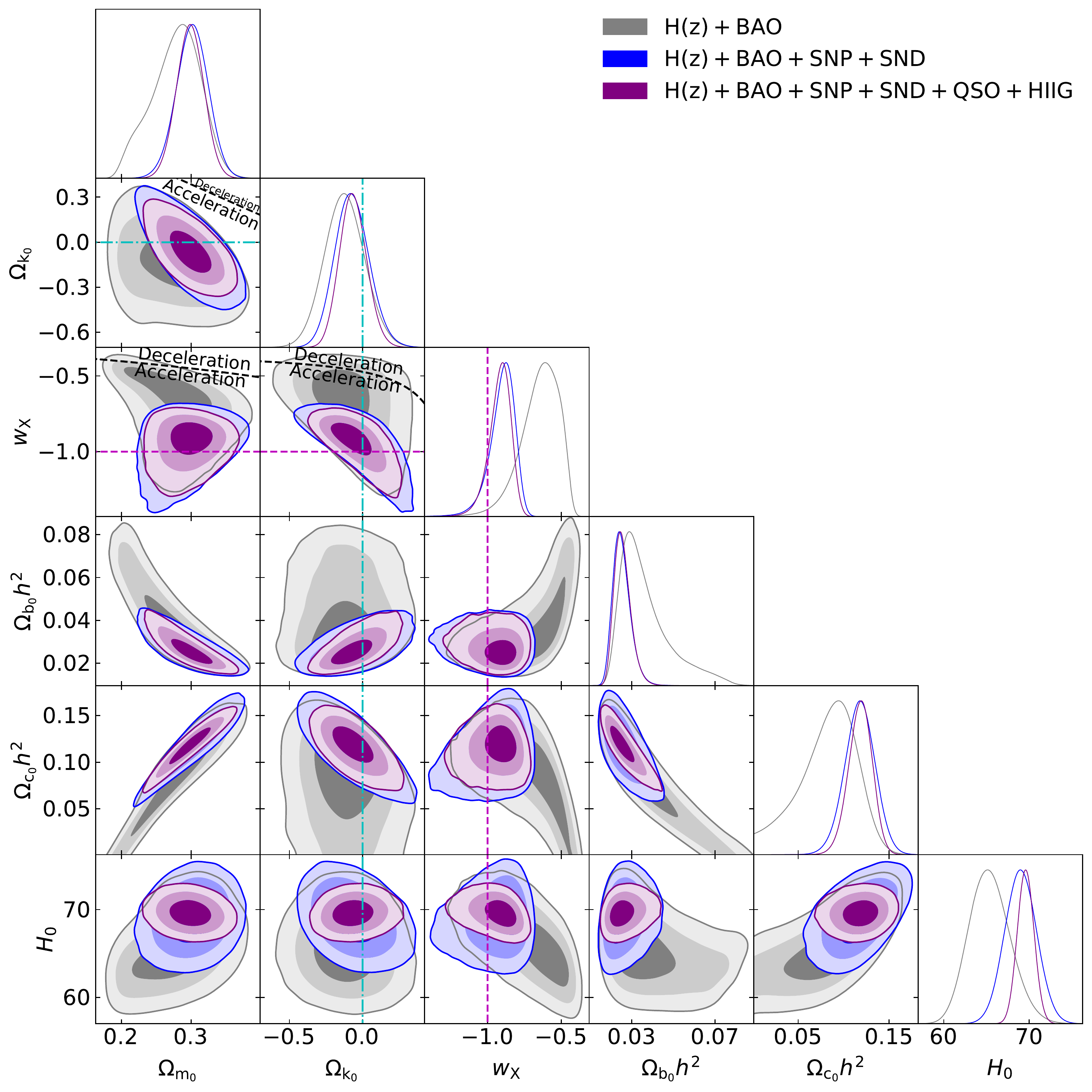}}\\
\caption{Same as Fig. \ref{fig3} but for non-flat XCDM, where the zero acceleration lines in each of the three subpanels of the right panel are computed for the third cosmological parameter set to the $H(z)$ + BAO data best-fitting values listed in Table \ref{tab:BFP}. Currently-accelerating cosmological expansion occurs below these lines. The cyan dash-dot lines represent the flat XCDM case, with closed spatial hypersurfaces either below or to the left. In all cases, almost all of the favored parameter space is associated with currently-accelerating cosmological expansion. The magenta lines indicate $w_{\rm X} = -1$, i.e. the non-flat \lcdm\ model.}
\label{fig4}
\end{figure*}

\begin{figure*}
\centering
 \subfloat[]{%
    \includegraphics[width=3.5in,height=3.5in]{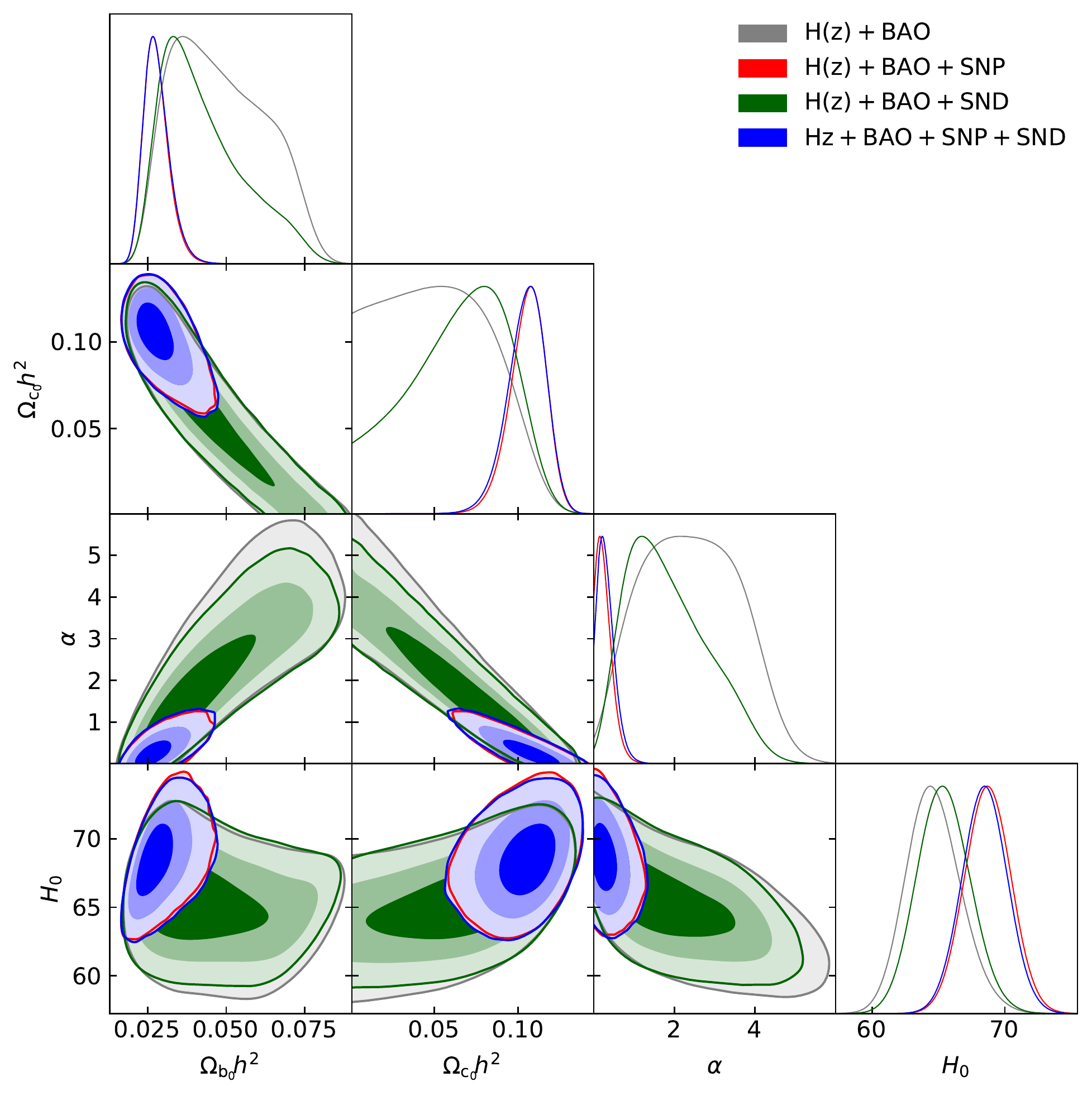}}
 \subfloat[]{%
    \includegraphics[width=3.5in,height=3.5in]{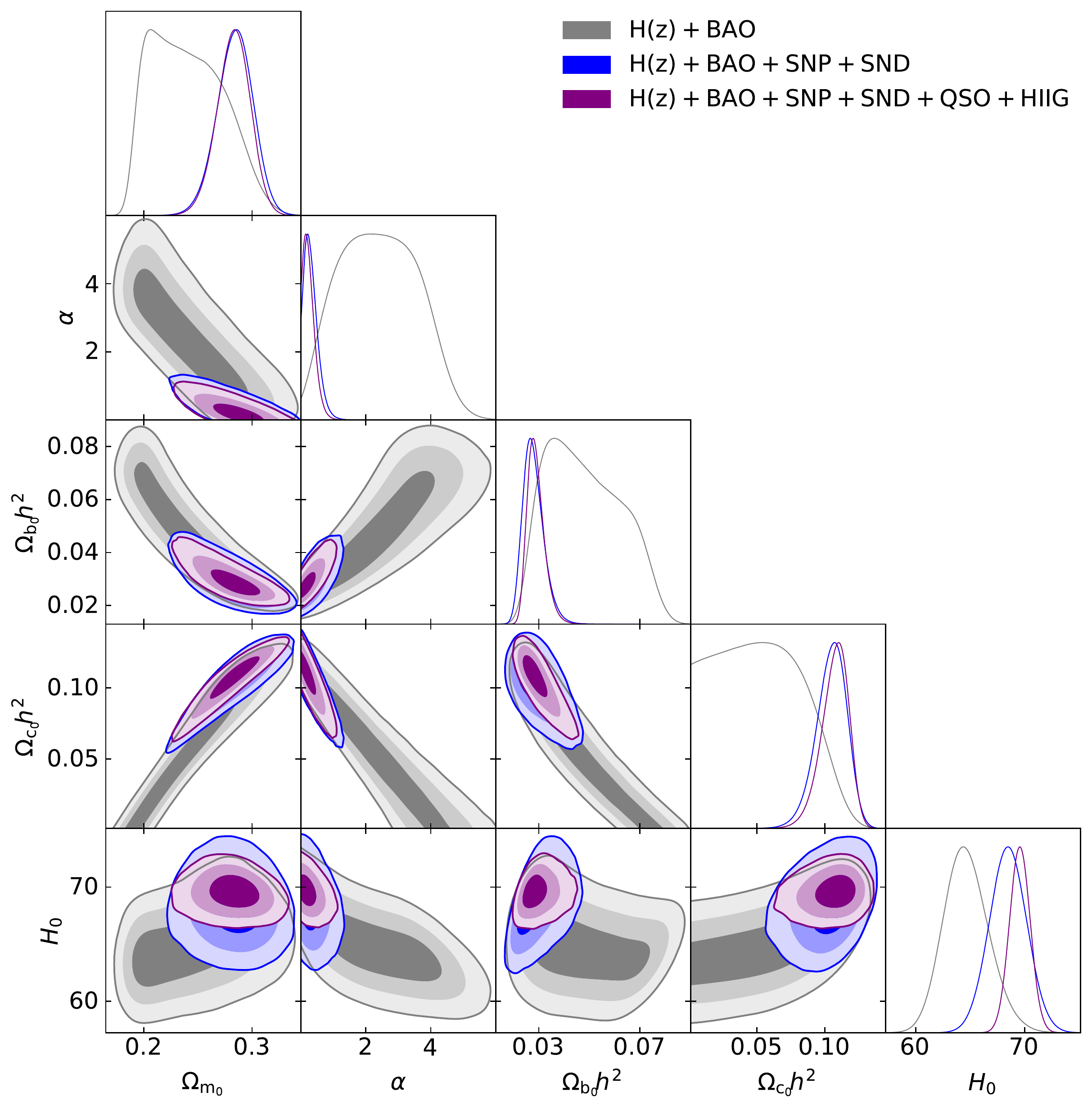}}\\
\caption{1$\sigma$, 2$\sigma$, and 3$\sigma$ confidence contours for flat \pcdm. In all cases, the favored parameter space is associated with currently-accelerating cosmological expansion. The $\alpha = 0$ axis is the flat \lcdm\ model.}
\label{fig5}
\end{figure*}

\begin{figure*}
\centering
 \subfloat[]{%
    \includegraphics[width=3.5in,height=3.5in]{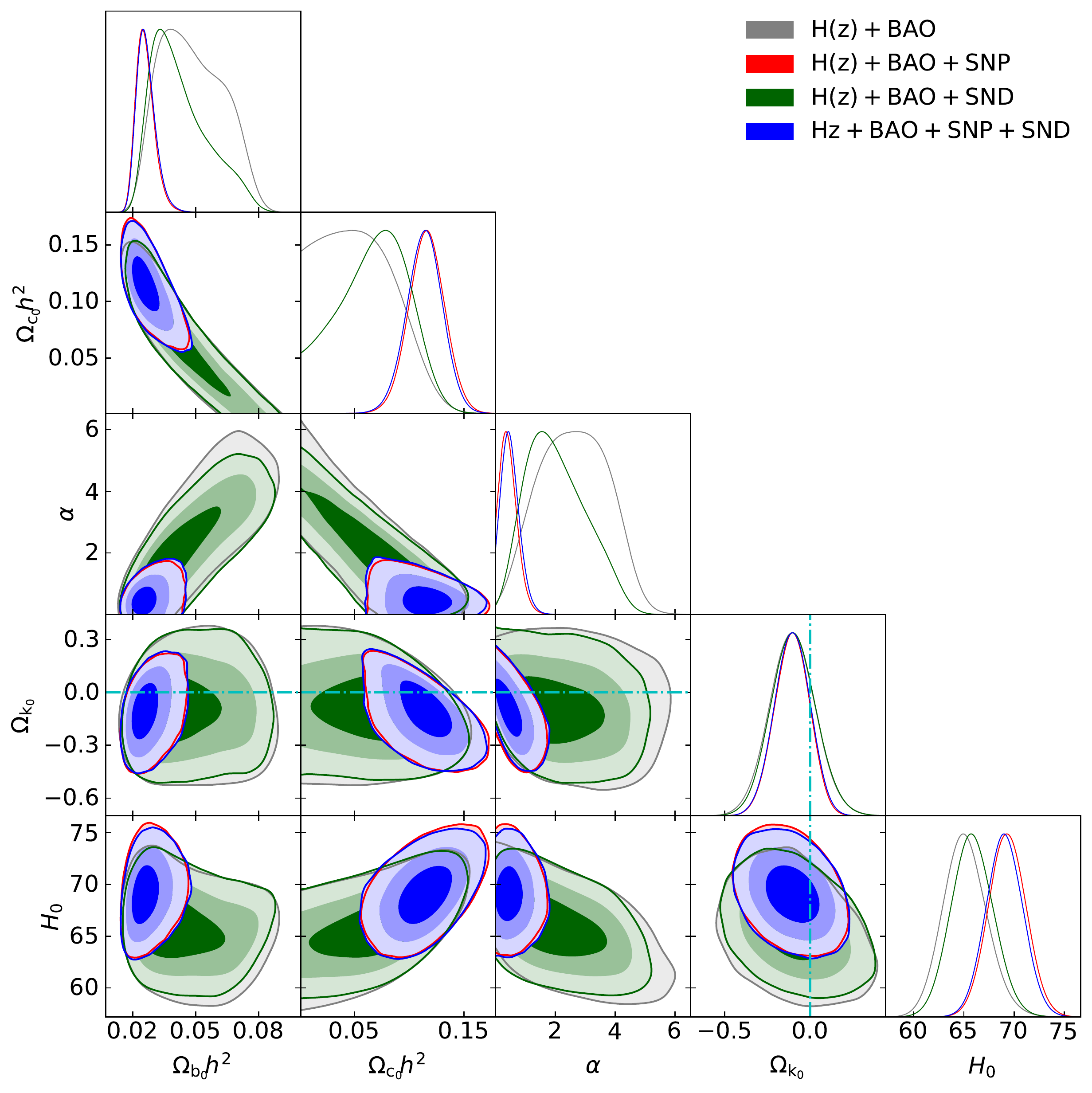}}
 \subfloat[]{%
    \includegraphics[width=3.5in,height=3.5in]{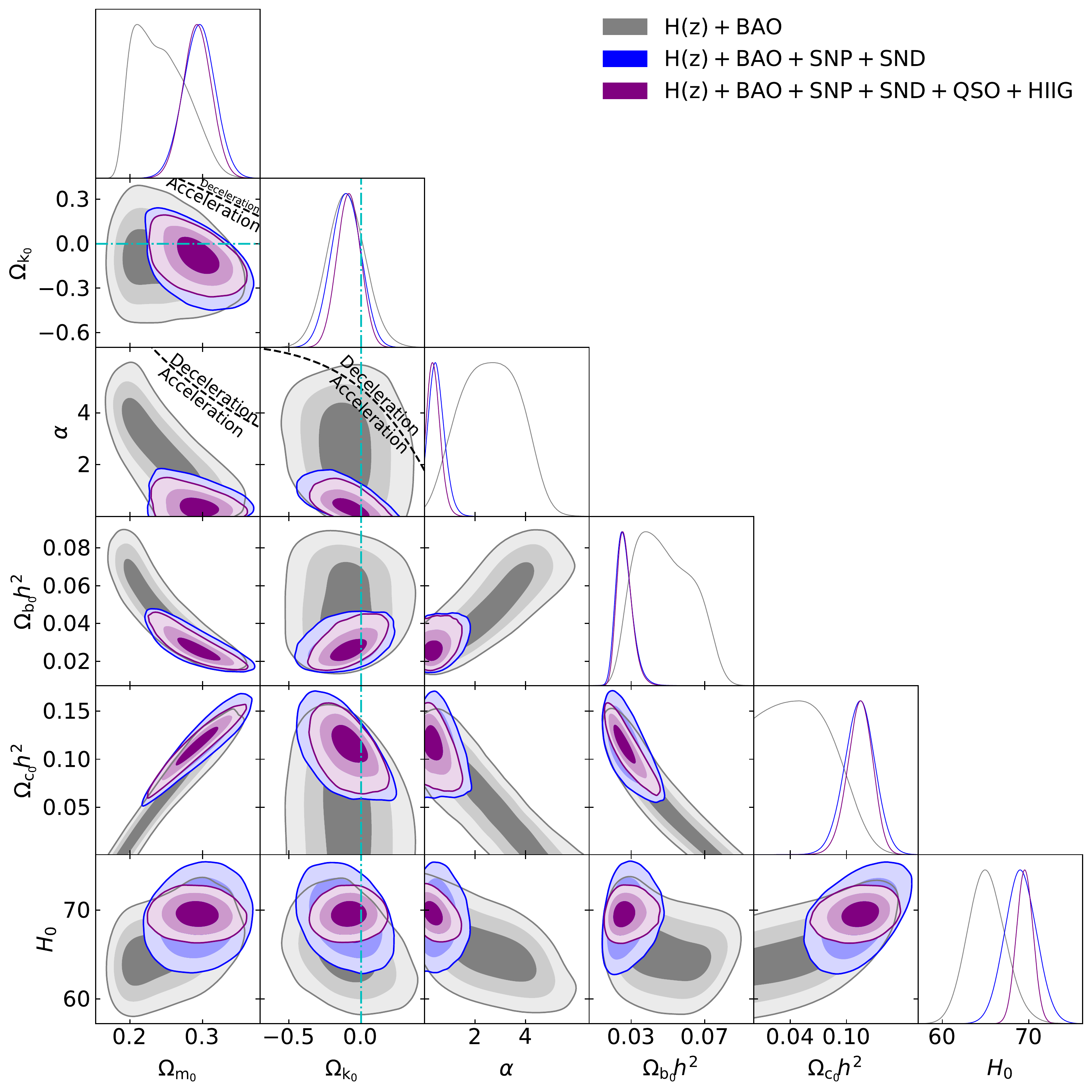}}\\
\caption{Same as Fig. \ref{fig5} but for non-flat \pcdm, where the zero-acceleration lines in each of the sub-panels of the right panel are computed for the third cosmological parameter set to the $H(z)$ + BAO data best-fitting values listed in Table \ref{tab:BFP}. Currently-accelerating cosmological expansion occurs below these lines. In all cases, almost all of the favored parameter space is associated with currently-accelerating cosmological expansion. The cyan dash-dot lines represent the flat \pcdm\ case, with closed spatial geometry either below or to the left. The $\alpha = 0$ axis is the non-flat \lcdm\ model.}
\label{fig6}
\end{figure*}

The posterior one-dimensional (1D) probability distributions and two-dimensional (2D) confidence regions of the cosmological parameters for the six flat and non-flat models are shown in Figs. \ref{fig1}--\ref{fig6}, in gray ($H(z)$+BAO), red ($H(z)$ + BAO + SN-Pantheon, HzBSNP), green ($H(z)$ + BAO + SN-DES, HzBSND), blue ($H(z)$ + BAO + SN-Pantheon + SN-DES, HzBSNPD), and purple ($H(z)$ + BAO + SN-Pantheon + SN-DES + QSO + \hiig, HzBSNPDQH). We list the unmarginalized best-fitting parameter values, as well as the corresponding $\chi^2$, $AIC$, $BIC$, and degrees of freedom $\nu$ ($\nu \equiv N - n$) for all models and data combinations, in Table \ref{tab:BFP}. The marginalized best-fitting parameter values and uncertainties ($\pm 1\sigma$ error bars or $2\sigma$ limits), for all models and data combinations, are listed in Table \ref{tab:1d_BFP}.\footnote{The \textsc{python} package \textsc{getdist} \citep{Lewis_2019} is used to analyze the samples.}

\subsection{$H(z)$ + BAO, HzBSNP, and HzBSND constraints}
\label{subsec:HzB}

The 1D marginalized $H(z)$ + BAO constraints on the cosmological parameters are listed in Table \ref{tab:1d_BFP}. These are (slightly) different from the ones obtained by \cite{Khadka_2020d}, because of the different treatments of both the prior ranges and the coefficient $\kappa$ in the \pcdm\ models.\footnote{We treated $\kappa$ as a derived constant determined from the parameter $\alpha$ (see e.g. eq. (14) of \citealp{Caoetal_2020b}), while \cite{Khadka_2020d} treated it as a constant derived from the energy budget equation.}

The $H(z)$, BAO, and SN-Pantheon data combinations have previously been studied \citep{park_ratra_2019b}. Relative to that analysis, we use the updated BAO data, shown in Table \ref{tab:BAO}, in our analysis here. In the HzBSNP case, we find that the determinations of \ok\ are more consistent with flat spatial hypersurfaces than what \cite{park_ratra_2019b} found and dark energy dynamics favors less deviation from a cosmological constant in the XCDM cases, while favoring a somewhat stronger deviation from $\alpha=0$ in the non-flat \pcdm\ case.

Because the $H(z)$, BAO, and SN-DES constraints are consistent across all six of the models we study, we also perform a joint analysis of these data to determine HzBSND constraints. Relative to the HzBSNP constraints, the measured values of $\Omega_{\rm b_0}\!h^2$, $\Omega_{\rm c_0}\!h^2$, and \om\ are a little higher, lower, and lower (except for flat \lcdm) than those values measured from the HzBSNP case, respectively. Given the error bars, these differences are not statistically significant. The measured values of $H_0$ are lower than those for the HzBSNP case. The non-flat XCDM and \pcdm\ models favor more and less closed geometry than in the HzBSNP case. The non-flat \lcdm\ model favors more open geometry than in the HzBSNP case. The constraints for all three non-flat models are consistent with spatially flat hypersurfaces. The fits to the HzBSND data produce stronger evidence for dark energy dynamics than the fits to the HzBSNP data.

\begin{table*}
\centering
\begin{threeparttable}
\caption{Unmarginalized best-fitting parameter values for all models from various combinations of data.}\label{tab:BFP}
\setlength{\tabcolsep}{1.5mm}{
\begin{tabular}{lcccccccccccc}
\toprule
Model & Data set & $\Omega_{\mathrm{b_0}}\!h^2$ & $\Omega_{\mathrm{c_0}}\!h^2$ & $\Omega_{\mathrm{m_0}}$ & $\Omega_{\mathrm{k_0}}$ & $w_{\mathrm{X}}$ & $\alpha$ & $H_0$\tnote{a} & $\chi^2$ & $\nu$ & $AIC$ & $BIC$ \\
\midrule
Flat \lcdm & $H(z)$ + BAO & 0.0240 & 0.1179 & 0.299 & -- & -- & -- & 69.11 & 23.64 & 39 & 29.64 & 34.86\\
 & HzBSNP\tnote{b} & 0.0240 & 0.1180 & 0.299 & -- & -- & -- & 69.10 & 1053.22 & 1087 & 1059.22 & 1074.21\\
 & HzBSND\tnote{c} & 0.0234 & 0.1203 & 0.305 & -- & -- & -- & 68.82 & 50.83 & 59 & 56.83 & 63.21\\
 & HzBSNPD\tnote{d} & 0.0236 & 0.1196 & 0.303 & -- & -- & -- & 68.91 & 1080.46 & 1107 & 1086.46 & 1101.50\\
 & HzBSNPDQH\tnote{e} & 0.0251 & 0.1203 & 0.299 & -- & -- & -- & 69.92 & 1844.99 & 1380 & 1850.99 & 1866.69\\
\\
Non-flat \lcdm & $H(z)$ + BAO & 0.0248 & 0.1136 & 0.294 & 0.026 & -- & -- & 68.75 & 23.58 & 38 & 31.58 & 38.53\\
 & HzBSNP\tnote{b} & 0.0241 & 0.1172 & 0.298 & 0.004 & -- & -- & 69.06 & 1053.22 & 1086 & 1061.22 & 1081.20\\
 & HzBSND\tnote{c} & 0.0258 & 0.1081 & 0.292 & 0.071 & -- & -- & 67.92 & 50.28 & 58 & 58.28 & 66.79\\
 & HzBSNPD\tnote{d} & 0.0245 & 0.1150 & 0.297 & 0.023 & -- & -- & 68.68 & 1080.35 & 1106 & 1088.35 & 1108.40\\
 & HzBSNPDQH\tnote{e} & 0.0249 & 0.1209 & 0.300 & $-0.004$ & -- & -- & 69.93 & 1844.99 & 1379 & 1852.99 & 1873.92\\
\\
Flat XCDM & $H(z)$ + BAO & 0.0323 & 0.0860 & 0.280 & -- & $-0.696$ & -- & 65.12 & 19.65 & 38 & 27.65 & 34.60\\
 & HzBSNP\tnote{b} & 0.0254 & 0.1120 & 0.292 & -- & $-0.951$ & -- & 68.72 & 1052.63 & 1086 & 1060.63 & 1080.61\\
 & HzBSND\tnote{c} & 0.0300 & 0.0934 & 0.286 & -- & $-0.752$ & -- & 65.90 & 45.46 & 58 & 53.46 & 61.97\\
 & HzBSNPD\tnote{d} & 0.0256 & 0.1107 & 0.293 & -- & $-0.932$ & -- & 68.43 & 1079.23 & 1106 & 1087.23 & 1107.28\\
 & HzBSNPDQH\tnote{e} & 0.0268 & 0.1136 & 0.291 & -- & $-0.949$ & -- & 69.63 & 1844.27 & 1379 & 1852.27 & 1873.20\\
\\
Non-flat XCDM & $H(z)$ + BAO & 0.0302 & 0.0956 & 0.294 & $-0.155$ & $-0.650$ & -- & 65.55 & 18.31 & 37 & 28.31 & 37.00\\
 & HzBSNP\tnote{b} & 0.0234 & 0.1231 & 0.307 & $-0.103$ & $-0.895$ & -- & 69.25 & 1051.82 & 1085 & 1061.82 & 1086.79\\
 & HzBSND\tnote{c} & 0.0277 & 0.1046 & 0.301 & $-0.136$ & $-0.711$ & -- & 66.45 & 44.34 & 57 & 54.34 & 64.98\\
 & HzBSNPD\tnote{d} & 0.0236 & 0.1220 & 0.307 & $-0.107$ & $-0.877$ & -- & 68.98 & 1078.36 & 1105 & 1088.36 & 1113.42\\
 & HzBSNPDQH\tnote{e} & 0.0242 & 0.1217 & 0.303 & $-0.092$ & $-0.900$ & -- & 69.54 & 1843.25 & 1378 & 1853.25 & 1879.41\\
\\
Flat $\phi$CDM & $H(z)$ + BAO & 0.0361 & 0.0758 & 0.264 & -- & -- & 1.484 & 65.30 & 19.48 & 38 & 27.48 & 34.43\\
 & HzBSNP\tnote{b} & 0.0260 & 0.1145 & 0.292 & -- & -- & 0.101 & 69.51 & 1051.46 & 1086 & 1059.46 & 1079.44\\
 & HzBSND\tnote{c} & 0.0328 & 0.0860 & 0.273 & -- & -- & 1.061 & 66.16 & 45.17 & 58 & 53.17 & 61.68\\
 & HzBSNPD\tnote{d} & 0.0254 & 0.1102 & 0.292 & -- & -- & 0.168 & 68.35 & 1078.18 & 1106 & 1086.18 & 1106.22\\
 & HzBSNPDQH\tnote{e} & 0.0264 & 0.1135 & 0.290 & -- & -- & 0.132 & 69.57 & 1842.95 & 1379 & 1850.95 & 1871.88\\
\\
Non-flat $\phi$CDM & $H(z)$ + BAO & 0.0354 & 0.0811 & 0.269 & $-0.148$ & -- & 1.819 & 66.06 & 18.16 & 37 & 28.16 & 36.85\\
 & HzBSNP\tnote{b} & 0.0234 & 0.1225 & 0.305 & $-0.133$ & -- & 0.393 & 69.32 & 1050.31 & 1085 & 1060.31 & 1085.28\\
 & HzBSND\tnote{c} & 0.0319 & 0.0933 & 0.282 & $-0.140$ & -- & 1.411 & 66.84 & 44.09 & 57 & 54.09 & 64.72\\
 & HzBSNPD\tnote{d} & 0.0256 & 0.1159 & 0.298 & $-0.080$ & -- & 0.377 & 69.09 & 1077.13 & 1105 & 1087.13 & 1112.19\\
 & HzBSNPDQH\tnote{e} & 0.0258 & 0.1155 & 0.293 & $-0.078$ & -- & 0.354 & 69.55 & 1842.00 & 1378 & 1852.00 & 1878.16\\
\bottomrule
\end{tabular}}
\begin{tablenotes}[flushleft]
\item [a] \hunit.
\item [b] $H(z)$ + BAO + SN-Pantheon.
\item [c] $H(z)$ + BAO + SN-DES.
\item [d] $H(z)$ + BAO + SN-Pantheon + SN-DES.
\item [e] $H(z)$ + BAO + SN-Pantheon + SN-DES + QSO + \hiig.
\end{tablenotes}
\end{threeparttable}
\end{table*}

\subsection{HzBSNPD constraints}
\label{subsec:HzBSNPD}

The results of the previous three subsections show that, when combined with $H(z)$ + BAO data, SN-Pantheon data produce tighter constraints on almost all cosmological parameters, than do SN-DES data (with a few exceptions including $\Omega_{\rm b_0}\!h^2$ for non-flat \lcdm, $\Omega_{\rm c_0}\!h^2$ for non-flat \pcdm, and \om\ and $H_0$ for flat and non-flat \pcdm). Since the $H(z)$ + BAO, SN-Pantheon, and SN-DES data constraints are not inconsistent, it is useful to derive constraints from an analysis of the combined $H(z)$, BAO, SN-Pantheon, and SN-DES (HzBSNPD) data. The results of such an analysis are presented in this subsection. We discuss these results in some detail here because, as discussed in Sec. \ref{subsec:comparison}, we believe that the constraints we obtain from the HzBSNPD data combination are more reliable than the constraints we obtain from the other data combinations we study.

The measured values of $\Omega_{\rm b_0}\!h^2$ range from a low of $0.0241^{+0.0024}_{-0.0030}$ (flat \lcdm) to a high of $0.0279^{+0.0031}_{-0.0048}$ (flat \pcdm) and those of $\Omega_{\rm c_0}\!h^2$ range from a low of $0.1047^{+0.0125}_{-0.0096}$ (flat \pcdm) to a high of $0.1199\pm0.0067$ (flat \lcdm). The derived constraints on \om\ range from a low of $0.284^{+0.017}_{-0.016}$ (flat \pcdm) to a high of $0.303\pm0.013$ (flat \lcdm). These measurements are consistent with what is measured by \cite{planck2018b}. In particular, for flat \lcdm, comparing to the TT,TE,EE+lowE+lensing results in Table 2 of \cite{planck2018b} the error bars we find here for \obhs\!, \ochs\!, and \om\ are a factor of 18, 5.6, and 1.8, respectively, larger than the \textit{Planck} error bars, and our estimates here for the quantities differ from the \textit{Planck} estimates by 0.58$\sigma$, 0.015$\sigma$, and 0.82$\sigma$, respectively.

The constraints on $H_0$ are between $H_0=68.48^{+1.71}_{-1.70}$ \hunit\ (flat \pcdm) and $H_0=69.14\pm1.68$ \hunit\ (flat \lcdm), which are $0.35\sigma$ (flat \lcdm) and $0.15\sigma$ (flat \pcdm) higher than the median statistics estimate of $H_0=68 \pm 2.8$ \hunit\ \citep{chenratmed}, and $2.22\sigma$ (flat \lcdm) and $2.50\sigma$ (flat \pcdm) lower than the local Hubble constant measurement of $H_0 = 74.03 \pm 1.42$ \hunit\ \citep{riess_etal_2019}.\footnote{Other local expansion rate $H_0$ measurements result in slightly lower central values with slightly larger error bars \citep{rigault_etal_2015,zhangetal2017,Dhawan,FernandezArenas,freedman_etal_2020,rameez_sarkar_2019,Breuvaletal_2020, Efstathiou_2020, Khetan_et_al_2020}. Our $H_0$ determinations are consistent with earlier median statistics determinations \citep{gott_etal_2001,chen_etal_2003} as well as with other recent $H_0$ measurements \citep{chen_etal_2017,DES_2018,Gomez-ValentAmendola2018, planck2018b,dominguez_etal_2019,Cuceu_2019,zeng_yan_2019,schoneberg_etal_2019,lin_ishak_2019, Blum_et_al_2020, Lyu_et_al_2020, Philcox_et_al_2020, Zhang_Huang_2020, Birrer_et_al_2020, Denzel_et_al_2020,Pogosianetal_2020,Boruahetal_2020,Kimetal_2020,Harvey_2020}.} For flat \lcdm\ our $H_0$ error bar is a factor of 3.1 larger than that from the \textit{Planck} data and our $H_0$ estimate is 1.01$\sigma$ higher than that of \textit{Planck}.

For non-flat \lcdm, non-flat XCDM, and non-flat \pcdm, we find $\Omega_{\rm k_0}=0.032\pm0.072$, $\Omega_{\rm k_0}=-0.071^{+0.110}_{-0.123}$, and $\Omega_{\rm k_0}=-0.105\pm0.104$, respectively, with non-flat \pcdm\ favoring closed geometry at 1.01$\sigma$. The non-flat XCDM and \pcdm\ models favor closed geometry, while the non-flat \lcdm\ model favors open geometry. The constraints for non-flat \lcdm\ and XCDM models are consistent with spatially flat hypersurfaces.

The fits to the HzBSNPD data favor dark energy dynamics, where for flat (non-flat) XCDM, $w_{\rm X}=-0.932\pm0.061$ ($w_{\rm X}=-0.904^{+0.098}_{-0.058}$), with best-fitting value being 1.11$\sigma$ (1.66$\sigma$) away from $w_{\rm X}=-1$; and for flat (non-flat) \pcdm, $\alpha=0.320^{+0.108}_{-0.277}$ ($\alpha=0.509^{+0.212}_{-0.370}$), with best-fitting value being 1.16$\sigma$ (1.38$\sigma$) away from $\alpha=0$.

\begin{table*}
\centering
\begin{threeparttable}
\caption{One-dimensional marginalized best-fitting parameter values and uncertainties ($\pm 1\sigma$ error bars or $2\sigma$ limits) for all models from various combinations of data.}\label{tab:1d_BFP}
\setlength{\tabcolsep}{0.7mm}{
\begin{tabular}{lcccccccc}
\toprule
Model & Data set & $\Omega_{\mathrm{b_0}}\!h^2$ & $\Omega_{\mathrm{c_0}}\!h^2$ & $\Omega_{\mathrm{m_0}}$ & $\Omega_{\mathrm{k_0}}$ & $w_{\mathrm{X}}$ & $\alpha$ & $H_0$\tnote{a}\\
\midrule
Flat \lcdm & $H(z)$ + BAO & $0.0245^{+0.0026}_{-0.0032}$ & $0.1182\pm0.0077$ & $0.298^{+0.015}_{-0.017}$ & -- & -- & -- & $69.33\pm1.75$ \\
 & HzBSNP\tnote{b} & $0.0245^{+0.0025}_{-0.0031}$ & $0.1182\pm0.0068$ & $0.298\pm0.013$ & -- & -- & -- & $69.32\pm1.70$ \\
 & HzBSND\tnote{c} & $0.0239^{+0.0025}_{-0.0032}$ & $0.1206\pm0.0076$ & $0.305^{+0.015}_{-0.017}$ & -- & -- & -- & $69.04\pm1.74$ \\
 & HzBSNPD\tnote{d} & $0.0241^{+0.0024}_{-0.0030}$ & $0.1199\pm0.0067$ & $0.303\pm0.013$ & -- & -- & -- & $69.14\pm1.68$ \\
 & HzBSNPDQH\tnote{e} & $0.0253^{+0.0019}_{-0.0022}$ & $0.1202\pm0.0057$ & $0.299\pm0.012$ & -- & -- & -- & $69.98\pm0.91$ \\
\\
Non-flat \lcdm & $H(z)$ + BAO & $0.0265^{+0.0035}_{-0.0059}$ & $0.1104\pm0.0192$ & $0.291\pm0.024$ & $0.047^{+0.095}_{-0.112}$ & -- & -- & $68.71\pm2.24$ \\
 & HzBSNP\tnote{b} & $0.0253^{+0.0033}_{-0.0049}$ & $0.1158^{+0.0161}_{-0.0160}$ & $0.296\pm0.022$ & $0.013\pm0.073$ & -- & -- & $69.22\pm1.86$ \\
 & HzBSND\tnote{c} & $0.0276^{+0.0038}_{-0.0062}$ & $0.1049^{+0.0188}_{-0.0187}$ & $0.288\pm0.024$ & $0.090^{+0.093}_{-0.106}$ & -- & -- & $67.92\pm2.10$ \\
 & HzBSNPD\tnote{d} & $0.0257^{+0.0033}_{-0.0050}$ & $0.1133\pm0.0160$ & $0.295\pm0.022$ & $0.032\pm0.072$ & -- & -- & $68.83\pm1.82$ \\
 & HzBSNPDQH\tnote{e} & $0.0260^{+0.0031}_{-0.0046}$ & $0.1188^{+0.0138}_{-0.0123}$ & $0.297\pm0.020$ & $0.007\pm0.063$ & -- & -- & $69.95\pm0.93$ \\
\\
Flat XCDM & $H(z)$ + BAO & $0.0372^{+0.0045}_{-0.0138}$ & $0.0777^{+0.0351}_{-0.0182}$ & $0.270^{+0.036}_{-0.022}$ & -- & $-0.688^{+0.174}_{-0.109}$ & -- & $65.22^{+2.21}_{-2.64}$ \\
 & HzBSNP\tnote{b} & $0.0261^{+0.0030}_{-0.0041}$ & $0.1118\pm0.0105$ & $0.292\pm0.016$ & -- & $-0.951\pm0.063$ & -- & $68.91\pm1.76$ \\
 & HzBSND\tnote{c} & $0.0331^{+0.0038}_{-0.0091}$ & $0.0881^{+0.0235}_{-0.0137}$ & $0.279^{+0.027}_{-0.019}$ & -- & $-0.739^{+0.110}_{-0.108}$ & -- & $65.95\pm2.08$ \\
 & HzBSNPD\tnote{d} & $0.0264^{+0.0031}_{-0.0042}$ & $0.1105\pm0.0107$ & $0.292\pm0.016$ & -- & $-0.932\pm0.061$ & -- & $68.62\pm1.73$ \\
 & HzBSNPDQH\tnote{e} & $0.0273^{+0.0026}_{-0.0035}$ & $0.1131^{+0.0104}_{-0.0095}$ & $0.291\pm0.015$ & -- & $-0.949\pm0.059$ & -- & $69.67^{+0.97}_{-0.96}$ \\
\\
Non-flat XCDM & $H(z)$ + BAO & $0.0367^{+0.0049}_{-0.0145}$ & $0.0822^{+0.0376}_{-0.0233}$ & $0.278^{+0.041}_{-0.030}$ & $-0.122^{+0.137}_{-0.136}$ & $-0.647^{+0.159}_{-0.084}$ & -- & $65.39^{+2.18}_{-2.59}$ \\
 & HzBSNP\tnote{b} & $0.0251^{+0.0031}_{-0.0049}$ & $0.1186\pm0.0167$ & $0.301\pm0.023$ & $-0.066^{+0.111}_{-0.124}$ & $-0.923^{+0.104}_{-0.060}$ & -- & $69.24\pm1.87$ \\
 & HzBSND\tnote{c} & $0.0315^{+0.0039}_{-0.0091}$ & $0.0956^{+0.0260}_{-0.0190}$ & $0.290^{+0.031}_{-0.026}$ & $-0.099\pm0.133$ & $-0.714^{+0.116}_{-0.089}$ & -- & $66.30\pm2.14$ \\
 & HzBSNPD\tnote{d} & $0.0253^{+0.0032}_{-0.0048}$ & $0.1178^{+0.0166}_{-0.0165}$ & $0.301\pm0.023$ & $-0.071^{+0.110}_{-0.123}$ & $-0.904^{+0.098}_{-0.058}$ & -- & $69.00\pm1.85$ \\
 & HzBSNPDQH\tnote{e} & $0.0256^{+0.0030}_{-0.0046}$ & $0.1182^{+0.0136}_{-0.0121}$ & $0.299\pm0.020$ & $-0.063^{+0.087}_{-0.097}$ & $-0.919^{+0.085}_{-0.056}$ & -- & $69.59\pm0.97$ \\
\\
Flat $\phi$CDM & $H(z)$ + BAO & $0.0480^{+0.0113}_{-0.0195}$ & $0.0524^{+0.0246}_{-0.0427}$ & $0.240^{+0.024}_{-0.044}$ & -- & -- & $2.418^{+1.197}_{-1.331}$ & $64.67^{+1.86}_{-2.22}$ \\
 & HzBSNP\tnote{b} & $0.0278^{+0.0030}_{-0.0046}$ & $0.1055^{+0.0119}_{-0.0091}$ & $0.284\pm0.016$ & -- & -- & $<0.666$ & $68.71^{+1.73}_{-1.74}$ \\
 & HzBSND\tnote{c} & $0.0429^{+0.0071}_{-0.0170}$ & $0.0641^{+0.0371}_{-0.0235}$ & $0.251^{+0.038}_{-0.031}$ & -- & -- & $1.863^{+0.674}_{-1.316}$ & $65.41^{+1.91}_{-2.08}$ \\
 & HzBSNPD\tnote{d} & $0.0279^{+0.0031}_{-0.0048}$ & $0.1047^{+0.0125}_{-0.0096}$ & $0.284^{+0.017}_{-0.016}$ & -- & -- & $0.320^{+0.108}_{-0.277}$ & $68.48^{+1.71}_{-1.70}$ \\
 & HzBSNPDQH\tnote{e} & $0.0289^{+0.0025}_{-0.0040}$ & $0.1073^{+0.0116}_{-0.0081}$ & $0.283^{+0.016}_{-0.014}$ & -- & -- & $0.261^{+0.067}_{-0.254}$ & $69.57\pm0.94$ \\
\\
Non-flat $\phi$CDM & $H(z)$ + BAO & $0.0482^{+0.0126}_{-0.0190}$ & $0.0544^{+0.0194}_{-0.0497}$ & $0.242^{+0.024}_{-0.046}$ & $-0.103\pm0.132$ & -- & $2.618^{+1.213}_{-1.226}$ & $65.14^{+2.02}_{-2.29}$\\
 & HzBSNP\tnote{b} & $0.0260^{+0.0033}_{-0.0051}$ & $0.1159^{+0.0163}_{-0.0161}$ & $0.296\pm0.022$ & $-0.106\pm0.102$ & -- & $0.454^{+0.174}_{-0.372}$ & $69.33\pm1.86$ \\
 & HzBSND\tnote{c} & $0.0427^{+0.0076}_{-0.0177}$ & $0.0670^{+0.0379}_{-0.0282}$ & $0.253^{+0.037}_{-0.039}$ & $-0.097\pm0.130$ & -- & $2.058^{+0.779}_{-1.269}$ & $65.86\pm2.09$ \\
 & HzBSNPD\tnote{d} & $0.0264^{+0.0034}_{-0.0052}$ & $0.1139\pm0.0161$ & $0.295\pm0.022$ & $-0.105\pm0.104$ & -- & $0.509^{+0.212}_{-0.370}$ & $69.06^{+1.84}_{-1.83}$ \\
 & HzBSNPDQH\tnote{e} & $0.0265^{+0.0031}_{-0.0048}$ & $0.1142^{+0.0141}_{-0.0123}$ & $0.293\pm0.020$ & $-0.085\pm0.081$ & -- & $0.399^{+0.159}_{-0.313}$ & $69.53\pm0.95$ \\
\bottomrule
\end{tabular}}
\begin{tablenotes}[flushleft]
\item [a] \hunit.
\item [b] $H(z)$ + BAO + SN-Pantheon.
\item [c] $H(z)$ + BAO + SN-DES.
\item [d] $H(z)$ + BAO + SN-Pantheon + SN-DES.
\item [e] $H(z)$ + BAO + SN-Pantheon + SN-DES + QSO + \hiig.
\end{tablenotes}
\end{threeparttable}
\end{table*}

\subsection{HzBSNPDQH constraints}
\label{subsec:HzBSNPDQH}
Since the constraints derived from $H(z)$, BAO, SN-Pantheon, SN-DES, QSO, and \hiig\ data are not inconsistent, in this subsection we jointly analyze HzBSNPDQH data to determine more restrictive constraints on the cosmological parameters (though as discussed in Sec. \ref{subsec:comparison}, we believe these constraints to be less reliable than those that stem from the HzSNPD combination, so we only describe the broad outlines here).

For flat \lcdm, the error bars we derive for \obhs\!, \ochs\!, and \om\ are larger than the \textit{Planck} error bars, though our central estimates of these quantities are broadly consistent with those derived from \textit{Planck}. In a similar fashion, we find larger error bars on $H_0$ in flat \lcdm\ than does \textit{Planck}, though our central estimate is higher than theirs. Generally, the constraints we derive on $H_0$ are more consistent with the median statistics estimate of $H_0=68 \pm 2.8$ \hunit\ \citep{chenratmed}, than with the local Hubble constant measurement of $H_0 = 74.03 \pm 1.42$ \hunit\ \citep{riess_etal_2019}.

We find mild evidence for spatial curvature, with non-flat XCDM and \pcdm\ favoring closed geometry, and non-flat \lcdm\ mildly favoring open geometry. The constraints from non-flat \lcdm\ and XCDM are consistent with spatially flat hypersurfaces to within less than 1$\sigma$. Additionally, we find mild evidence for dark energy dynamics, with the best-fitting value of $w_{\rm X}$ being 0.86$\sigma$ (1.45$\sigma$) away from $w_{\rm X}=-1$ in flat (non-flat) XCDM, and the best-fitting value of $\alpha$ being 1.03$\sigma$ (1.27$\sigma$) away from $\alpha=0$ in flat (non-flat) \pcdm.

\subsection{Model comparison}
\label{subsec:comparison}

The values of $\Delta\chi^2$, $\Delta AIC$, $\Delta BIC$, and the reduced $\chi^2$ ($\chi^2/\nu$) are reported in Table \ref{tab:cab}, where $\Delta \chi^2$, $\Delta AIC$, and $\Delta BIC$, respectively, are defined as the differences between the values of the $\chi^2$, $AIC$, and $BIC$ for a given model and their corresponding minimum values among all models. From Table \ref{tab:cab}, we see that the reduced $\chi^2$ values determined from the $H(z)$ + BAO data combination range from 0.49 to 0.62, which is probably due to the $H(z)$ data having overestimated error bars (see \citealp{Caoetal_2020b} for discussions of the systematic errors of these data). As discussed in \cite{Ryan_2} and \cite{Caoetal_2020a}, the underestimated systematic uncertainties in QSO and \hiig\ data\footnote{Roberto Terlevich and his colleagues are currently investigating the systematic uncertainties of the \hiig\ data, the results of which they plan to publish in a future paper (Roberto Terlevich, private communication, 2021).} result in larger reduced $\chi^2$ ($\sim1.34$) for the models in the HzBSNPDQH case. The reduced $\chi^2$ values for the HzBSNP and HzBSNPD cases are around unity for all models and for the HzBSND case range from 0.77 to 0.87. Of the combinations we study here, on the basis of these reduced $\chi^2$ values, the HzBSNPD constraints should be viewed as the most reliable ones.

We find that based on the $AIC$ and $BIC$, flat \lcdm\ and flat \pcdm\ are the most favored models in different data combination cases. The $\Delta AIC$ results show that the most favored model is flat \lcdm\ in the HzBSNP case, while the most favored model is flat \pcdm\ in the rest of the data combinations. The $\Delta BIC$ results show that the most favored model is flat \pcdm\ in the $H(z)$ + BAO and HzBSND cases, and is flat \lcdm\ in the remaining cases. For both $\Delta AIC$ and $\Delta BIC$ results, the most disfavored model is non-flat \lcdm\ in the $H(z)$ + BAO and HzBSND cases, and is non-flat XCDM in all other cases, with positive evidence against non-flat \lcdm\ and either positive or very strong evidence (depending on the data combination) against non-flat XCDM.

Overall, the $\Delta AIC$ results show no strong evidence against any model, and neither do the $\Delta BIC$ results for the $H(z)$ + BAO and HzBSND cases. However, in the HzBSNP and HzBSNPDQH cases, the $\Delta BIC$ results show strong evidence against the non-flat \lcdm\ and flat XCDM models, and very strong evidence against the non-flat \pcdm\ and XCDM models. In the HzBSNPD case, the evidence against flat XCDM and flat \pcdm\ is positive, the evidence against non-flat \lcdm\ is strong, and the evidence against non-flat \pcdm\ and non-flat XCDM is very strong. Based on the $\Delta \chi^2$ results, non-flat \pcdm\ has the minimum $\chi^2$ in all cases. 

In summary, the HzBSNPD data favor flat \pcdm\ ($AIC$) or flat \lcdm\ ($BIC$) among the six models we study here.

\begin{table*}
\centering
\begin{threeparttable}
\caption{$\Delta \chi^2$, $\Delta AIC$, $\Delta BIC$, and $\chi^2_{\mathrm{min}}/\nu$ values.}\label{tab:cab}
\setlength{\tabcolsep}{2.0mm}{
\begin{tabular}{lccccccc}
\toprule
 Quantity & Data set & Flat \lcdm & Non-flat \lcdm & Flat XCDM & Non-flat XCDM & Flat \pcdm & Non-flat \pcdm\\
\hline
 & $H(z)$ + BAO & 5.48 & 5.42 & 1.49 & 0.15 & 1.32 & 0.00\\
 & HzBSNP\tnote{a} & 2.91 & 2.91 & 2.32 & 1.51 & 1.15 & 0.00 \\
$\Delta \chi^2$ & HzBSND\tnote{b} & 6.74 & 6.19 & 1.37 & 0.25 & 1.08 & 0.00\\
 & HzBSNPD\tnote{c} & 3.33 & 3.22 & 2.10 & 1.23 & 1.05 & 0.00\\
 & HzBSNPDQH\tnote{d} & 2.99 & 2.99 & 2.27 & 1.25 & 0.95 & 0.00\\
 \\
 & $H(z)$ + BAO & 2.16 & 4.10 & 0.17 & 0.83 & 0.00 & 0.68\\
 & HzBSNP\tnote{a} & 0.00 & 2.00 & 1.41 & 2.60 & 0.24 & 1.09 \\
$\Delta AIC$ & HzBSND\tnote{b} & 3.66 & 5.11 & 0.29 & 1.17 & 0.00 & 0.92\\
 & HzBSNPD\tnote{c} & 0.28 & 2.17 & 1.05 & 2.18 & 0.00 & 0.95\\
 & HzBSNPDQH\tnote{d} & 0.04 & 2.04 & 1.32 & 2.30 & 0.00 & 1.05\\
 \\
 & $H(z)$ + BAO & 0.43 & 4.10 & 0.17 & 2.57 & 0.00 & 2.42\\
 & HzBSNP\tnote{a} & 0.00 & 6.99 & 6.40 & 12.58 & 5.23 & 11.07 \\
$\Delta BIC$ & HzBSND\tnote{b} & 1.53 & 5.11 & 0.29 & 3.30 & 0.00 & 3.04\\
 & HzBSNPD\tnote{c} & 0.00 & 6.90 & 5.78 & 11.92 & 4.72 & 10.69\\
 & HzBSNPDQH\tnote{d} & 0.00 & 7.23 & 6.51 & 12.72 & 5.19 & 11.47\\
 \\
 & $H(z)$ + BAO & 0.61 & 0.62 & 0.52 & 0.49 & 0.51 & 0.49\\
 & HzBSNP\tnote{a} & 0.97 & 0.97 & 0.97 & 0.97 & 0.97 & 0.97 \\
$\chi^2_{\mathrm{min}}/\nu$ & HzBSND\tnote{b} & 0.86 & 0.87 & 0.78 & 0.78 & 0.78 & 0.77\\
 & HzBSNPD\tnote{c} & 0.98 & 0.98 & 0.98 & 0.98 & 0.97 & 0.97\\
 & HzBSNPDQH\tnote{d} & 1.34 & 1.34 & 1.34 & 1.34 & 1.34 & 1.34\\
\bottomrule
\end{tabular}}
\begin{tablenotes}[flushleft]
\item [a] $H(z)$ + BAO + SN-Pantheon.
\item [b] $H(z)$ + BAO + SN-DES.
\item [c] $H(z)$ + BAO + SN-Pantheon + SN-DES.
\item [d] $H(z)$ + BAO + SN-Pantheon + SN-DES + QSO + \hiig.
\end{tablenotes}
\end{threeparttable}
\end{table*}

\section{Conclusion}
\label{sec:conclusion}

By analyzing a total of 1383 measurements, consisting of 31 $H(z)$, 11 BAO, 1048 SN-Pantheon, 20 SN-DES, 120 QSO, and 153 \hiig\ data points, we jointly constrain cosmological parameters in six flat and non-flat cosmological models.

From the constraints derived using the cosmological models, we can identify some relatively model-independent features. As discussed in Sec. \ref{subsec:comparison}, the $H(z)$ + BAO + SN-Pantheon + SN-DES (HzBSNPD) data combination produces the most reliable constraints. In particular, for the HzBSNPD data combination, we find a reasonable and fairly restrictive summary value of $\Omega_{\rm m_0}=0.294 \pm 0.020$,\footnote{Here we take the summary central value to be the mean of the two of six central-most values. As for the uncertainty, we call the difference between the two central-most values twice the systematic uncertainty and the average of the two central-most error bars the statistical uncertainty, and compute the summary error bar as the quadrature sum of the two uncertainties.} which is in good agreement with many other recent measurements (e.g. $0.315\pm0.007$ from \citealp{planck2018b}). A fairly restrictive summary value of $H_0=68.8\pm1.8$ \hunit\ is found to be in better agreement with the estimates of \cite{chenratmed} and \cite{planck2018b} than with the measurement of \cite{riess_etal_2019}; note that the constraints from BAO data do not depend on physics of the early Universe (with $\Omega_{\rm b_0}\!h^2$ being a free parameter that is fitted to the data used here). There is some room for dark energy dynamics or a little spatial curvature energy density in the HzBSNPD constraints, but based on $AIC$ and $BIC$ criteria, flat \pcdm\ or flat \lcdm\ are the best candidate models.

\section*{Acknowledgements}

We thank Javier de Cruz P\'{e}rez for useful discussions on the data. This work was partially funded by Department of Energy grant DE-SC0011840. The computing for this project was performed on the Beocat Research Cluster at Kansas State University, which is funded in part by NSF grants CNS-1006860, EPS-1006860, EPS-0919443, ACI-1440548, CHE-1726332, and NIH P20GM113109.

\section*{Data availability}

The \hiig\ data used in this article were provided to us by the authors of \cite{G-M_2019}. These data will be shared on request to the corresponding author with the permission of the authors of \cite{G-M_2019}.




\bibliographystyle{mnras}
\bibliography{mybibfile} 







\bsp	
\label{lastpage}
\end{document}